\documentclass{aa}  

\usepackage{graphicx}

\usepackage{threeparttable}
\usepackage{wasysym}
\usepackage{hyperref}

\usepackage{txfonts}
\usepackage{color}
\usepackage{float}
\usepackage{placeins}

\begin{document} 

   \title{Effect of stellar spot-only spectral lines in high-resolution transit spectroscopy}

   \author{W. Dethier
          \inst{1}
          \and
          E. Cristo
          \inst{1}
          \and
          O. Demangeon
          \inst{1}
          \and
          J. P. Lucero
          \inst{1,2}
          \and
          N. C. Santos
          \inst{1,2}
          \and
          B. Tessore
          \inst{3}
          }

   \institute{Instituto de Astrofísica e Ciências do Espaço, Universidade do Porto, CAUP, Rua das Estrelas, 4150-762 Porto, Portugal\\
              \email{william.dethier@astro.up.pt}
         \and
             Departamento de Física e Astronomia, Faculdade de Ciências, Universidade do Porto, Rua do Campo Alegre, 4169-007 Porto, Portugal
             \and
             Rue Menon, 38000, Grenoble, France
             }

   \date{}
  \abstract
   {High-resolution transit spectroscopy is a key technique for characterising exoplanet atmospheres. It relies on computing the absorption spectrum, which is sensitive to spectral discrepancies between the local spectra (absorbed by the planet) and the disc-integrated stellar spectra. While centre-to-limb variations are known contributors to these spectral differences, the impact of localised stellar features, such as spots containing different spectral information than the photosphere, remains inconclusive and under-explored.
   }
   {We investigate how a representative single spectral line present only in a stellar spot can introduce distortions in high-resolution planetary absorption spectra. Specifically, we assess whether such spot-induced features may resemble planetary atmospheric absorption and the observational or stellar conditions under which they become significant.}
   {Using the \texttt{SOAPv4} code, we simulated transit observations of an atmosphere-less planet across a star with a spot containing an additional absorption line compared to the photosphere's spectrum. Two systems were modelled: a close-in Jupiter-sized planet and a distant Earth-sized planet transiting a Sun-sized star. We varied the stellar rotation period and the spot size, and ensured that the planet transits the spot. Spectral lines were idealised Gaussians to isolate the impact of the spot-specific feature.
   }
   {In the close-in Jupiter-sized case, the spot-only line induces a pseudo-emission distortion in the mean in-transit absorption spectrum of a few tenths of a percent that does not resemble atmospheric absorption. However, in the distant Earth-sized case, the spot-only line could produce spurious absorption features of an order of magnitude comparable to expected atmospheric absorption signals. The origin of these distortions is not primarily the spot crossing event itself, but rather the discrepancy between in- and out-of-transit spectra caused by the spot motion. We found that the effect is more intense for long-period planets around faster-rotating stars. The size of the distortion scales with the stellar rotation and relative spot size.}
   {We show that a single stellar spot line can generate spurious features in the absorption spectrum. These features contaminate the absorption spectrum in ways that may lead to misinterpretation if not accounted for. For Jupiter-sized planets with short transits, the features are mainly emission-like, but can nonetheless dampen a potential true atmospheric absorption signature. For Earth-sized planets with long transits, the distortion is mainly absorption-like and could be interpreted as planetary absorption. These results highlight the importance of accounting for stellar heterogeneity and activity in the analysis of high-resolution absorption spectra.}


   \maketitle
   
%

\section{Introduction}
Transmission spectroscopy observations are one of the best current ways to study and characterise transiting exoplanet atmospheres. However, these observations are inherently affected by stellar spectral lines to the point that the standard technique to extract the absorption signature of a transiting planetary atmosphere does not get rid of these stellar contamination, but, on the contrary, induces distortions that complicate the accurate characterisation of the planetary atmospheres. These induced distortions originate from the non-similarity between the local stellar spectrum absorbed by the planet and the disc-integrated spectrum, both affected by centre-to-limb variations (CLVs)\footnote{In this study, when we refer to CLVs, we mean variations in the shape, the intensity level, and the wavelength's position of the line profile. The phenomena leading to these variations are: different stellar atmosphere depths are scanned by the observer's line-of-sight as a function of the limb-angle, limb darkening, and stellar rotation, respectively.} of the stellar line profiles \citep{louden2015, czesla2015, yan2017, casasayasbarris2020, casasayasbarris2021,borsa2021, dethier2023,carteret2024}. \\
To mitigate these distortions, it is common to use synthetic absorption spectra of an atmosphere-less planet to correct the absorption spectrum built from observations \citep[e.g.][]{Casasayasbarris2017, casasayasbarris2018, yan2018,nugroho2020,maguire2023,keles2024,maguire2024}. However, this approach still leaves biases that affect the atmospheric absorption signature \citep{dethier2024}. \\
These obstacles to the accurate characterisation of planetary atmospheres are already present for quiet homogeneous stars, so, naturally, one can wonder whether these distortions could be amplified by stellar activity or not \citep{cauley2017,cauley2018,lucero2026}, as it can locally change the spectrum of a star. More specifically, it is known that stellar spots, due to their cooler temperature compared to the surrounding photosphere, can contain absorption lines that are absent from the photosphere's spectrum or lines that are deeper, relative to the continuum, than in the photosphere's spectrum \citep{gray2005}. For example, absorption lines from neutral atoms such as the \ion{Li}{I} multiplet at 6708 $\AA$, some \ion{Fe}{I} lines and the \ion{Ti}{I} multiplet lines at 2.2 $\mu$m \citep[e.g.][]{ritzenhoff1997,Rueedi1998,solanki2003,smitha2021}, or from molecules that would dissociate in the hotter photosphere, with notable examples including the detections of TiO and VO \citep{vogt1979}, SiO \citep{glenar1983, campbell1995}, OH \citep{oneal1997}, and H$_2$O absorption lines in stellar spots \citep{wallace1995, wallace1996,sonnabend2006}. \\ 

Several studies have investigated the impact of occulted and unocculted stellar activity features on the transmission spectra of transiting planets. While there is broad agreement that stellar activity can influence transmission spectra, there is still no consensus on the magnitude of this effect or on the conditions under which it becomes significant  \citep{deming2013,mccullough2014,aronson2015,bruno2018,murgas2018,cauley2018,rackham2018,barclay2021,genest2022}. Notable examples include \citet{oshagh2014}, who showed that stellar facula occultation could mimic planetary atmospheric scattering for HD\,189733b and GJ\,3470b, and \citet{murgas2018}, who found a slope and excess absorption in the transmission spectrum of HAT-P-11b, which they attribute to unocculted spots. \citet{moran2023} who, based on low-resolution observations with the James Webb Space Telescope \citep{Gardner2006}, showed that the presence of water in spots could explain an observed excess absorption in the transmission spectrum of GJ\,486b. However, other works such as \citet{aronson2015} and \citet{bruno2018} mention that spots are unlikely to have a strong impact on the transmission spectra. 
Overall, while activity-induced contamination is well characterised for low-resolution transit spectroscopy \citep[e.g.][]{rackham2018}, its quantitative impact in high-resolution transit spectroscopy, where individual line distortions become significant, still remains less thoroughly explored \citep[e.g.][]{cauley2018,lucero2026}.\\

The motivation for this work is to gain a deeper comprehension of how stellar spot lines can impact high-resolution transit spectroscopy observations. Since distortions in absorption spectra of transiting planets originate from individual stellar lines, we aim to assess whether a single line present in a stellar spot, but absent from the quiet photosphere, can introduce such distortions. We also aim to evaluate whether these spot-induced line distortions are significant and if they could mimic a transiting atmosphere’s signal, potentially leading to false detections. Another goal is to understand the origin and behaviour of these distortions, and to qualitatively describe their shape and amplitude. 

To address these questions, we used a planetary transit simulation tool and explored a simplified case in which the spectrum of the stellar spot contains an additional single line not found in the quiet-photosphere's spectrum. We studied the effect of the presence of this line on the absorption spectrum of two different transiting planets without atmosphere: a close-in Jupiter-sized planet and a distant Earth-sized planet. Different configurations were tested to determine the conditions under which a spot-only line might introduce contaminating signatures.

In this study, we adopt a generalised theoretical approach to isolate a mechanism that arises when a spectral line is present in the spectrum of a stellar spot but absent from the photosphere's spectrum. We use idealised line profiles and simplified stellar models to characterise this effect independent of specific atomic or molecular transitions, stellar types, or wavelength regimes. References to particular systems (e.g. Earth-Sun, HD 189733b) serve only as illustrative benchmarks for parameter choices and do not constitute claims about those systems. 

In Sect. \ref{sec:spectrum}, we explain how we construct the spectrum for the photosphere and for the spot. In Sect. \ref{sec:simulation}, we detail the different simulations that we perform. In Sect. \ref{sec:result}, we show the results from simulations and discuss them. In Sect. \ref{sec:spotsize}, we explore the impact of spots with sizes not proportional to the planet's radius. We then conclude in Sect. \ref{sec:conclusion}. We also provide results from extra simulations done for specific cases in the Appendices, namely for different starting positions of the spot, different spot-only line depths, different temperature contrasts between the photosphere and the spot, and finally for a different wavelength range.

\section{The stellar photosphere and spot spectra}
\label{sec:spectrum}
\begin{figure}
    \centering    
    \includegraphics[width=\linewidth]{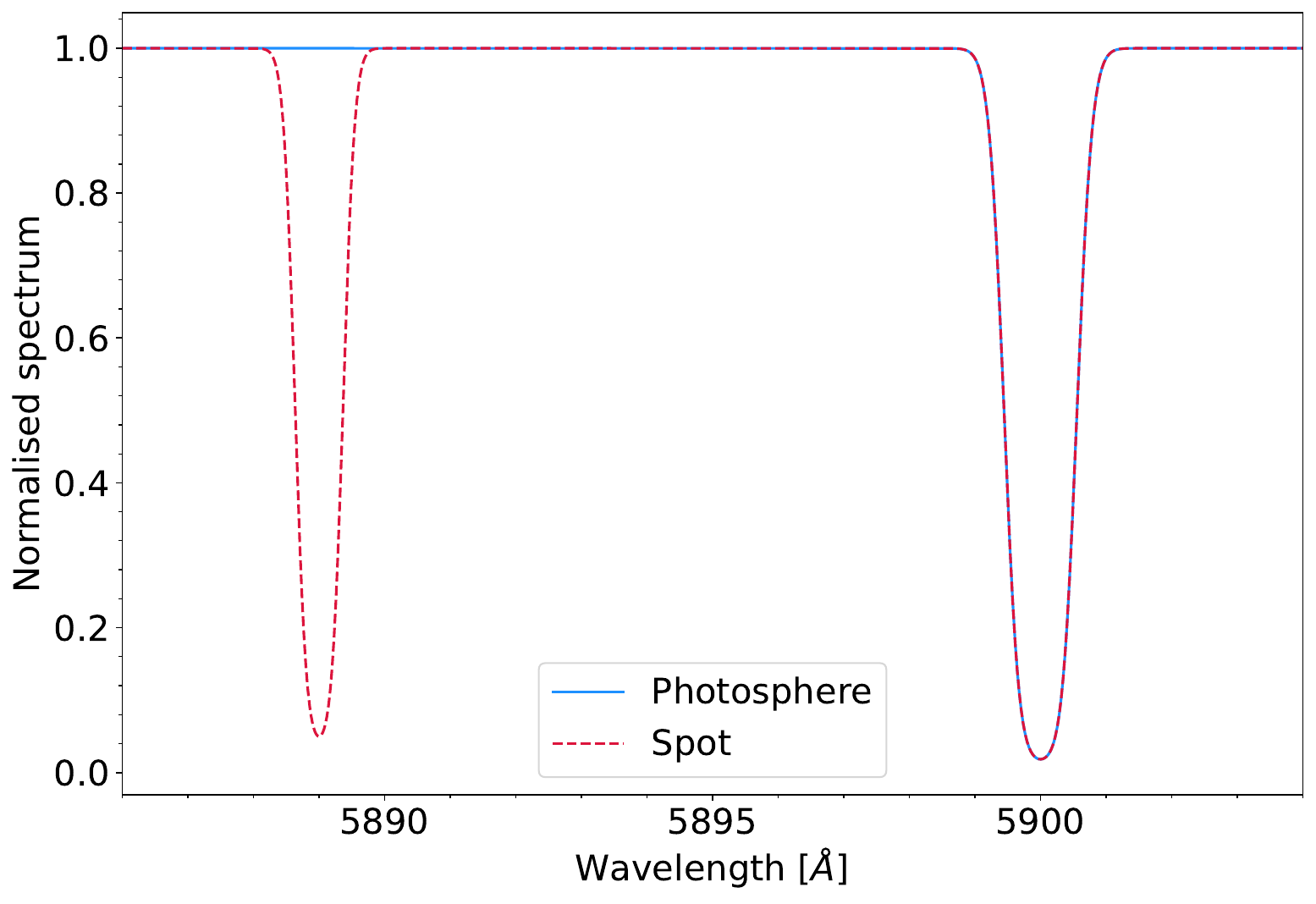}
    \caption{Input local spectra used in SOAPv4 for the spot and the photosphere. The effect on the spot's flux due to the temperature contrast is not represented in the figure, but is taken into account when building the stellar grid.}
    \label{fig:spectra}
\end{figure}

To construct the local spectra of the photosphere and of the spot, we adopted a purely hypothetical approach and simply modelled two absorption lines based on a Gaussian profile. Besides, as we aimed to isolate the effect of this spot-only line, we did not attempt to reproduce realistic line shapes, nor did the lines represent any specific transition. Even though the wavelength range covered the \ion{Na}{I} doublet, it was also chosen arbitrarily and used only as an illustrative wavelength range. Given the stellar effects considered in this study (see below), it should not influence the conclusions (see Appendix \ref{appendix:wavelength}). One of the lines was present in both the photosphere and in the spot spectra (at around 5900 \AA). The other one was only present in the spectrum of the spot (at around 5889 \AA). We also chose the depth and width of the lines to be of the same order of magnitude as the depth and width of the Solar \ion{Na}{I} D2 line, to provide a point of reference \citep{wallace1998}. Figure \ref{fig:spectra} shows the two normalised local spectra as a function of the chosen wavelength range\footnote{The wavelength step used was 0.01 \AA \ and no instrumental profile convolution was performed. We warn the reader that with a coarser wavelength sampling or by convolving the spectra with an instrumental profile, the exact shape of the distortions would change.}. We included the photospheric line in the simulations mainly for a reference visual comparison and do not enter into more details about the interpretation of the signal it induces, as our main focus is the spot-only line (for an in-depth analysis of the effect of spots on photospheric lines' distortions in transit absorption spectra, see \citet{lucero2026}). 

To build the stellar grid\footnote{The stellar disc was discretised into a square grid of 1000 by 1000 and 3000 by 3000 cells (diameter by diameter) for the Jupiter-sized and Earth-sized systems, respectively. These sizes minimise the numerical noise of grid discretisation, while allowing the simulations to remain fast.}, we used the fourth version of the code Spot Oscillation and Planet (SOAPv4) \citep{cristo2025}. In Figure \ref{fig:spectra}, the two spectra are normalised, but SOAPv4 takes as an argument the temperature contrast between the photosphere and the spot, which we set at 600K, to be on the order of magnitude of previous studies \citep{meunier2010,cristo2025}. Additionally, we used the same line profile for the whole quiet photosphere's spectrum; there was thus no variation in the shape of line profiles as a function of position on the stellar disc. The only CLVs that we accounted for were the wavelength shift of the local line profiles and the local flux variation caused by stellar rotation and limb darkening, respectively. The limb darkening was parameterised, and follows a quadratic law, with the quadratic coefficients $c_1 = 0.54292061$ and $c_2 = 0.13076036$ computed for the system's parameters ($\rm T_{eff}= 5772K$, log$(g) = 4.4374$ and metallicity = 0) over the simulated wavelength region with LDtK \citep{parviainen2015} and PHOENIX spectral library \citep{husser2013}.\\
This choice of restricting the spectral effects to stellar rotation and limb-darkening was motivated by our goal to better understand the origin and behaviour of the distortion produced by the spot-only line itself. Stellar rotation is the main contributor to planet-occulted line distortions (POLDs) \citep{casasayasbarris2021,dethier2023}. As for limb darkening, although it has a second-order effect on the POLDs \citep{dethier2023}, we still accounted for it, as the associated flux variations give more or less weight to the spectrum of the spot depending on its position. \\
We are aware that this simplified scenario neglects several physical effects such as CLV in the line shape, Zeeman and pressure broadening, umbra and penumbra structures, and plages. However, accounting for these effects or using more realistic spectra would introduce additional second-order effects that could combine with the spot contribution in complex, non-linear ways, making the interpretation of the results less straightforward. 

\section{Setup of the simulations}
\label{sec:simulation}
To simulate the planetary transit, we used the SOAPv4 code. We simulated two different systems, one with a close-in Jupiter-sized planet with the same planet-to-star radius ratio as HD\,189733 b \citep{bouchy2005}, a well-known hot-Jupiter system, and the other one with a long orbital period, Earth-sized planet with the same planet-to-star radius ratio as the Earth-Sun system. In both systems, the stellar radius was the radius of the Sun. For these two systems, we simulated different cases in which we varied the size of the spot with respect to the planetary radius and the stellar rotation period.\\
We used a stellar spot radius ($\rm S_p$) of 0.75, 1.0, and 1.25 times the planet's radius. For all the simulations, except in Appendix \ref{app:position}, the position of the spot on the apparent stellar disc at the middle of the transit (T$_0$) had a longitude = 30$^\circ$ and a latitude = 0$^\circ$, which ensured the planet crossed the spot during the transit. For the stellar rotation period (P$_*$), we used 5, 10 and 25 days, which correspond to approximately a $\rm v_{eq}$$\sin(i_*)=$ 10, 5 and 2 km s$^{-1}$, respectively, which cover typical values for slow-to-moderate rotators \citep[e.g.][]{nielsen2013}. Table \ref{tab:tab1} lists all the parameters of the system.\\
The SOAPv4 code delivered a time series of disc-integrated spectra in the stellar rest frame for all the exposures of the simulations\footnote{Similarly to what is done with observations, the code fits the out-of-transit points of the RV curve of the system with a linear curve to extract the local modulation caused by the motion of the planet and also, in our case, by the spot. All the individual spectra are then shifted in wavelength to the supposed stellar rest frame using these RV values.}, which we used to build the absorption spectrum. This was done by first normalising each disc-integrated spectrum to its continuum \citep[see e.g. ][]{casasayasbarris2020}. Then, we built the master out-of-transit spectrum $(\rm F_{out})$ by averaging all the normalised disc-integrated spectra from exposures outside of the transit's first and fourth contact times. Then, we computed, for each exposure, the difference between the master out-of-transit spectrum and the individual disc-integrated spectra. Finally, this difference was divided by the master out-of-transit spectrum to obtain the absorption spectra for each exposure. \\
For all the simulations, we made sure to have 1 hour before and after the transit to build the master out-of-transit spectrum. A one-hour baseline before and after transit is typically aimed for in transit observations because it provides an out-of-transit baseline comparable in duration to the transit. For the Earth-sized case, we also chose 1 hour, even though the transit duration is approximately 13 hours. This was intended to roughly mimic what is feasible from ground-based observations. This choice ensured that we had, for all the different simulations, a master out-of-transit spectrum computed over the same amount of time independently of the value of the rotation period of the star to highlight the impact of the latter on the results. This implies that $\rm F_{out}$ varies more with shorter stellar rotation periods and with longer transit. This is an effect we highlight and explore in this work. Moreover, we set the time step between exposures to be constant throughout all the cases. We also made sure that, in every case, several exposures happened when the planet passes in front of the spot. 

\begin{table}
\renewcommand{\arraystretch}{1.}
\setlength{\tabcolsep}{4pt}  
\caption{Stellar and planetary properties. }
\label{tab:tab1}
\centering
\begin{threeparttable}
\begin{tabular}{l c c c}
\hline\hline
\multicolumn{1}{l}{Parameter} & & \multicolumn{2}{c}{Value} \\
 & & Earth-sized & Jupiter-sized\\ 
\hline
\\
Stellar rotation period & (days) & \multicolumn{2}{c}{5, 10, 25} \\
Planet radius & ($\rm R_{\odot}$) & 0.0091 & 0.15\\
Planet mass & ($\rm M_{\oplus}$) & 1.0 & 317.8\\
Semi-major axis & ($\rm R_{\odot}$) & 215 & 8.85\\
Spot radius & ($\rm R_p$) & \multicolumn{2}{c}{0.75, 1.0, 1.25}\\
Spot longitude & (deg) & \multicolumn{2}{c}{30}\\
Spot latitude & (deg) & \multicolumn{2}{c}{0}\\
Spot temperature contrast & (K) & \multicolumn{2}{c}{600}\\
Orbital inclination & (deg) & \multicolumn{2}{c}{90}\\
Sky-projected spin-orbit& (deg) & \multicolumn{2}{c}{0}\\
angle&  & & \\
Eccentricity & & \multicolumn{2}{c}{0}\\
\hline                                  
\end{tabular}
\begin{tablenotes}
\item[] \tiny{\textbf{Note.} The positions of the spot are for T$_0$. The stellar rotation periods of 5, 10 and 25 days correspond, in our study, to a $\rm v_{eq}\sin(i_*)$ of 10, 5 and 2 km s$^{-1}$, respectively. } 
\end{tablenotes}
\end{threeparttable}
\end{table}

\section{Results and interpretation}
\label{sec:result}

\subsection{Close-in Jupiter-sized system}
\label{sec:jupiter}
\begin{figure}
    \centering
    \includegraphics[width=\linewidth]{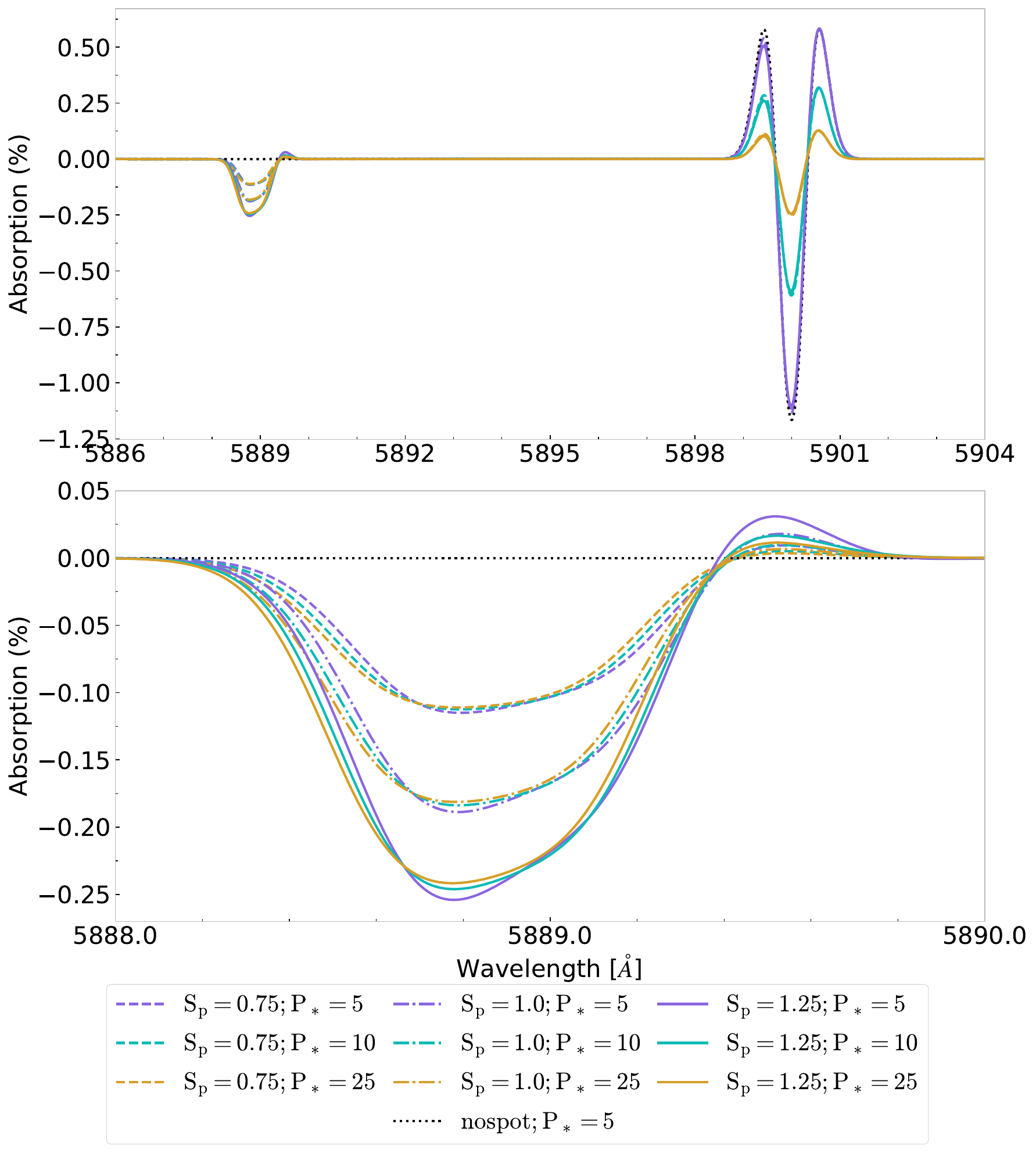}
    \caption{Upper panel: Absorption spectrum averaged between T$_2$ and T$_3$ in the planet's rest frame for the different cases for the Jupiter-sized system. Lower panel: Zoom on the wavelength range around the spot spectral line.}
    \label{fig:jupiter_abs_spec}
\end{figure}

\begin{figure}
    \centering
    \includegraphics[width=\linewidth]{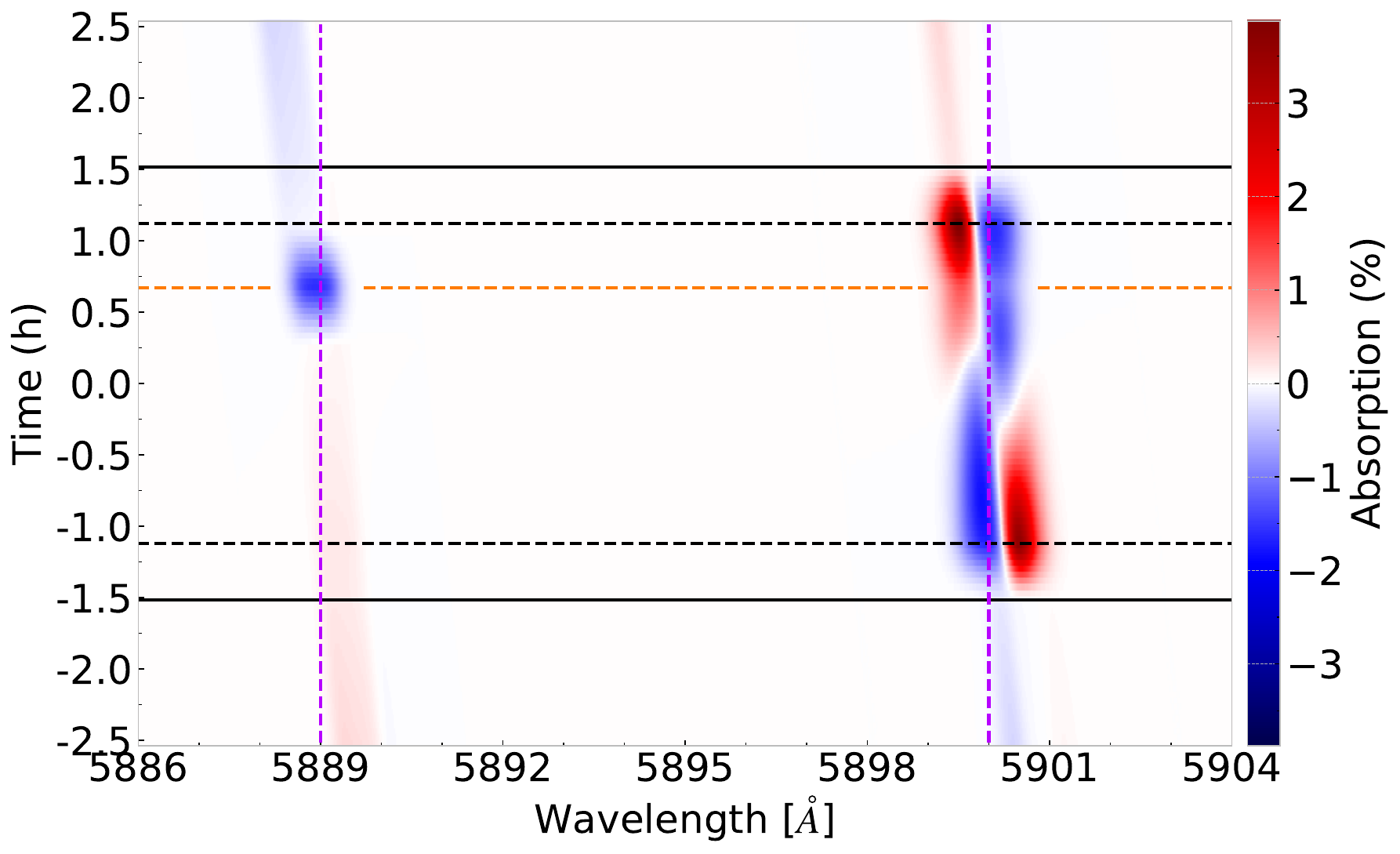}
    \caption{Absorption spectrum as a function of time and wavelength in the planet's rest frame for the case with $\rm s_p = 1.25$ and $\rm P_* = 5$ for a Jupiter-sized system. The horizontal plain and dashed black lines mark the contact times of the transit. The vertical dashed purple lines mark the reference wavelength of the stellar lines. The horizontal dashed orange line marks the spot crossing's time step.}
    \label{fig:jupiter_abs_spec_2D_sp125_P5}
\end{figure}

Figure \ref{fig:jupiter_abs_spec} shows the mean absorption spectrum in the planet's rest frame between T$_2$ and T$_3$\footnote{T$_2$ and T$_3$ are the second and third contact times of the transit, i.e. the times just after ingress and just before egress, respectively, e.g. \citet{seager2003}.} for the close-in Jupiter-sized system. \\
For the different cases, we recover the typical $w$ shape of the POLDs (mainly due to stellar rotation) for the strong photosphere line and an amplitude on the same order of magnitude as the POLDs expected for similar systems such as HD\,189733 and HD\,209458 (i.e. a few tenths of a percent to $\sim$ 1 $\%$ \citep{casasayasbarris2021,dethier2023}). As also expected, the amplitude of the POLDs for the photosphere line increases with $\rm v_{eq}$$\sin(i_*)$ \citep{dethier2023}. In the figures, as mentioned earlier, we show the signature around the photospheric line mainly for a visual comparison and do not enter into more details about the interpretation of the signal there. \\
For the spot-only line, there is also a POLD, though of lesser amplitude (a few tenths of a percent). In every case, the signal is emission-like (i.e. negative absorption in our figures) and could thus not mimic a planetary atmosphere absorption signature. Given the amplitude of atmospheric absorption signatures of some elements in typical close-in hot-Jupiters \citep[e.g.][]{wyttenbach2015,wyttenbach2017, chen2020, Grant2023}, an emission-like POLD  of a few tenths of a percent could dampen an actual absorption signature from the planet's atmosphere and lead to underestimates of the actual signature or even to false non-detections. The amplitude of the POLD increases with decreasing stellar rotation period, but the dominant cause is the relative size of the spot with respect to the size of the planet. \\
Figure \ref{fig:jupiter_abs_spec_2D_sp125_P5} shows the temporal evolution of the absorption spectrum in the planet's rest frame during the transit for the Jupiter-sized case with a spot size of 1.25 times the planetary radius and a stellar period of 5 days. We notice, around the spot-only line, that before the spot crossing (at a time $\approx$ 0.7 hours), the absorption spectrum is in absorption. Then, during the spot crossing and after, the absorption spectrum is in emission. In this case, we see that the spot crossing event dominates and we retrieve that signature in the mean absorption spectrum in Fig. \ref{fig:jupiter_abs_spec}. 

\begin{figure}
    \centering
    \includegraphics[width=\linewidth]{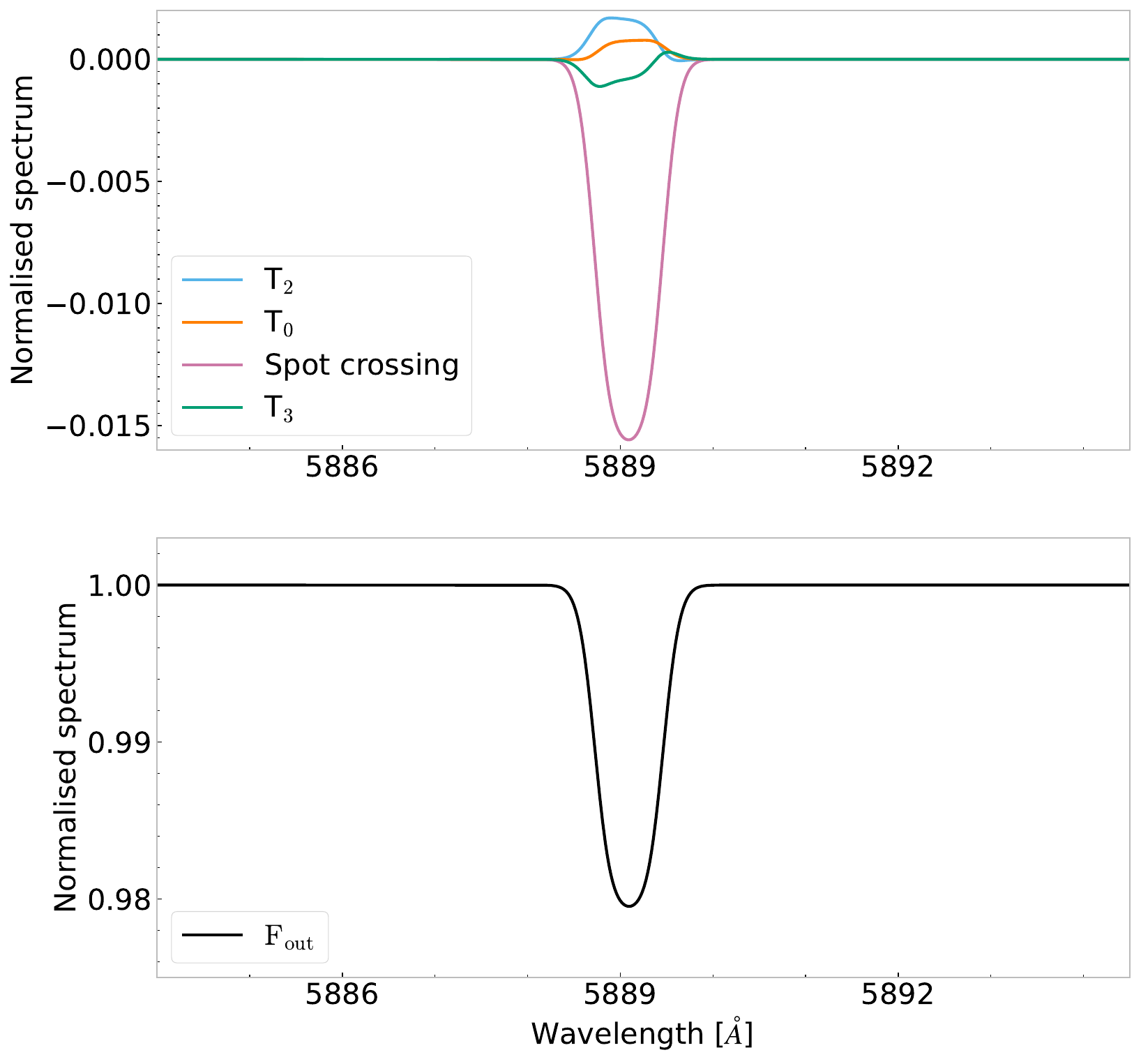}
    \caption{ Upper panel: Absorbed spectra at different symbolic time-steps as a function of wavelength for the case with $\rm s_p = 1.25$ and $\rm P_* = 5$ for the Jupiter-sized case. Lower panel: Normalised out-of-transit spectrum as a function of wavelength for the case with $\rm s_p = 1.25$ and $\rm P_* = 5$.}
    \label{fig:jupiter_absorbed_spec}
\end{figure}

Figure \ref{fig:jupiter_absorbed_spec} shows the out-of-transit spectrum and absorbed spectra ($\rm F_{abs} = F_{out} - F_{in}$) at different symbolic timesteps\footnote{Just after the ingress (T$_2$), middle of the transit (T$_0$), spot crossing and just before egress (T$_3$).} as a function of wavelength for the Jupiter-sized case. For the non spot-crossing time steps, the signal in $\rm F_{abs}$ is much smaller than for the time step of spot crossing. However, we do not expect to see the spot lines in the absorbed spectrum when the planet occults the quiet photosphere. The signal obs around the spot-only line in $\rm F_{abs}$ for those timesteps comes from the discrepancy between $\rm F_{out}$ and $\rm F_{in}$. Although they both contain a small\footnote{These lines are small because the spot contribution is averaged out by the disc-integration.} line at this wavelength position, the profiles of those lines are different. Indeed, the spot contribution is different in $\rm F_{out}$ and $\rm F_{in}$ due to the motion of the spot, and thus to changes in projected surface velocity, and lower fluxes (and consequently due to variations in the projected area of the spot). Therefore, when the difference is computed between $\rm F_{out}$ and $\rm F_{in}$, as the profile of the spot-only line is different, a signature appears.\\  
For the time step of the spot crossing, the signal is more intense than the other time steps because the spot-only line is actually present in the spectrum of the spot. The line in $\rm F_{out}$ is rather shallow, as in this region the disc-integrated spectrum contains only a small contribution from the spot. The results presented in Fig. \ref{fig:jupiter_abs_spec_2D_sp125_P5} are thus almost directly tracing what is observed in $\rm F_{abs}$.

\subsection{Earth-sized system on an Earth-like orbit}
\label{sec:earthfullsection}
In this section, we first show the results for the simulations using the full baseline to compute the master out-of-transit spectrum. Then, we show the results for the simulations using a master out-of-transit spectrum computed with only half of the baseline.
\subsubsection{Master out-of-transit spectrum computed with the full baseline}
\label{sec:earth}
\begin{figure}
    \centering
    \includegraphics[width=\linewidth]{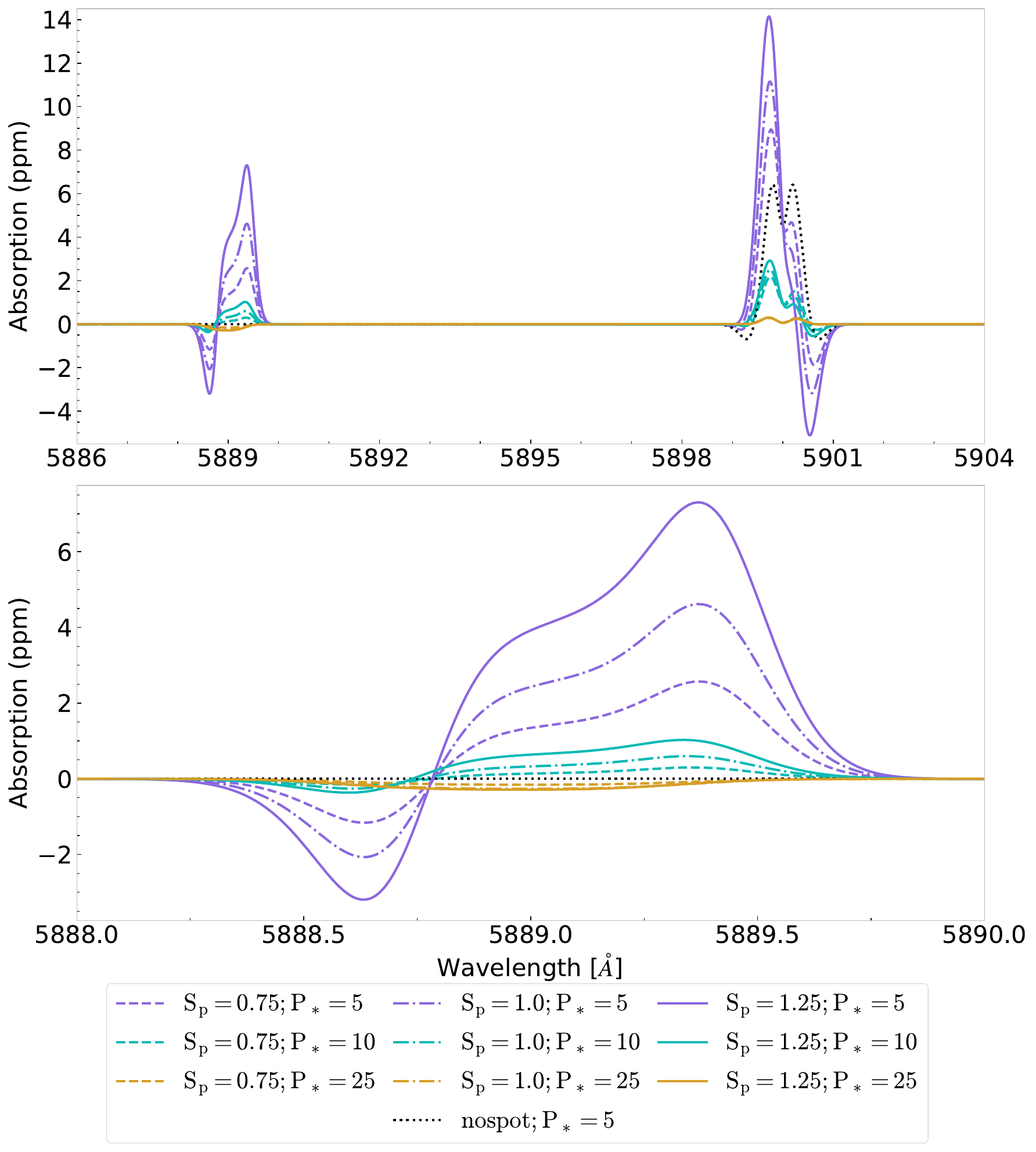}
    \caption{Upper panel: Absorption spectrum averaged between T$_2$ and T$_3$ in the planet's rest frame for the different cases for the Earth-sized system. Lower panel: Zoom on the wavelength range around the spot spectral line.}
    \label{fig:earth_abs_spec}
\end{figure}

\begin{figure}
    \centering
    \includegraphics[width=\linewidth]{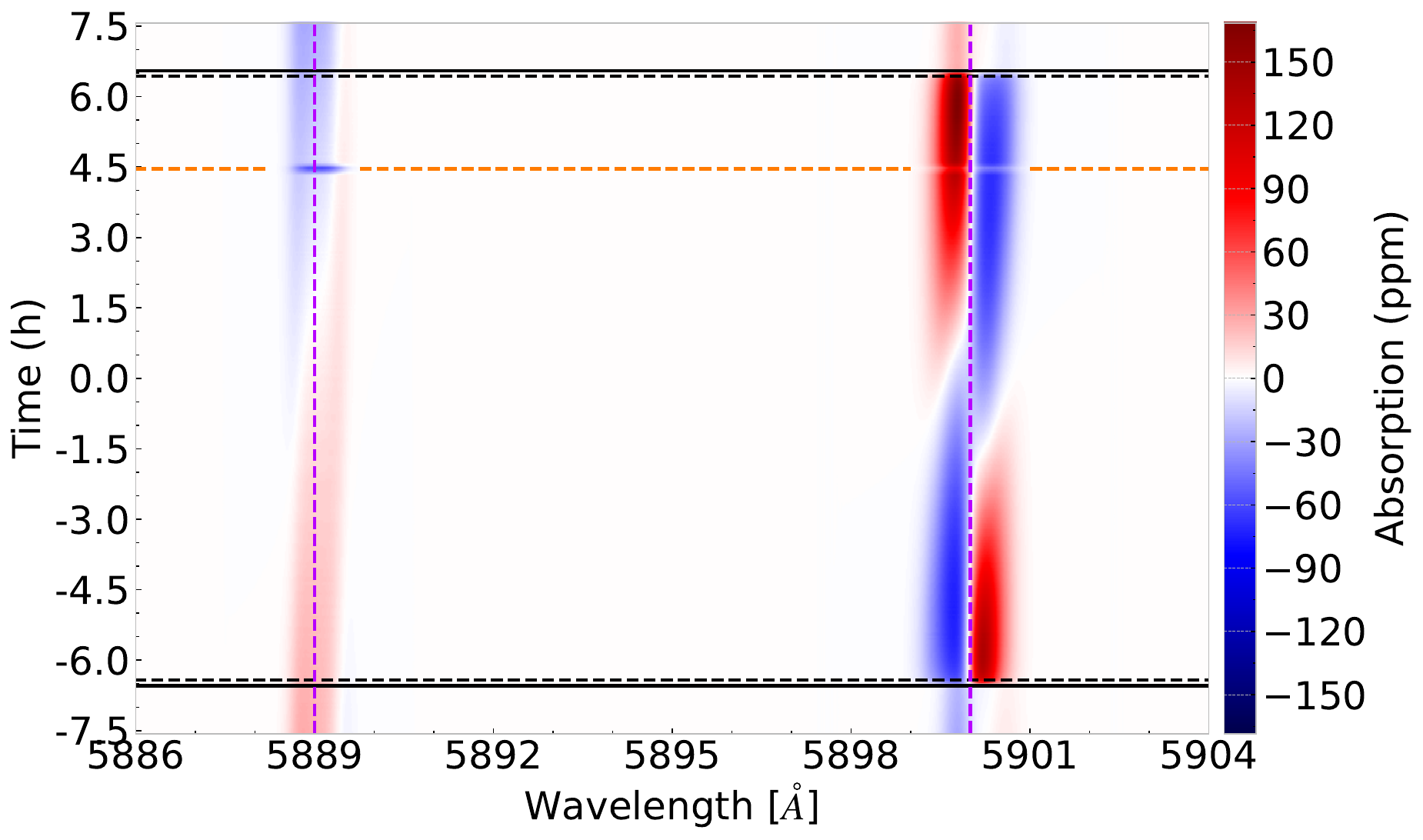}
    \caption{Absorption spectrum as a function of time and wavelength in the planet's rest frame for the case with $\rm s_p = 1.25$ and $\rm P_* = 5$ for an Earth-sized system. The horizontal plain and dashed black lines mark the contact times of the transit. The vertical dashed purple lines mark the reference wavelength of the stellar lines. The horizontal dashed orange line marks the spot crossing's time step.}
    \label{fig:earth_abs_spec_2D_sp125_P5}
\end{figure}

Figure \ref{fig:earth_abs_spec} shows the mean absorption spectrum in the planet's rest frame between T$_2$ and T$_3$ for the Earth-sized system. We retrieve large POLDs in the photosphere line only for cases with P$_* <= 10$ days, and the shape is quite different from the POLDs in the Jupiter-sized case. Similarly to the Jupiter-sized case, the amplitude of the photospheric line's POLDs is dominantly modulated by P$_*$. The amplitude increases as well with the size of the spot, which follows the expected trends as shown in \citet{lucero2026}. \\
For the spot-only line, a POLD is created and has an absorption-like signature for most cases. This mimics the signature of an absorbing atmosphere. The signal is dominantly modulated by P$_*$, but is also strongly impacted by the spot size, peaking from around 2 to 7 ppm for spot radii going from 0.75 to 1.25 times planetary radii, respectively. These amplitudes are of the expected order of magnitude of the amplitude of an Earth-Sun system's atmosphere signature \citep[i.e. a few parts per million (ppm),][]{ehrenreich2006,Ardaseva2017,Kreidberg2017}\\
Figure \ref{fig:earth_abs_spec_2D_sp125_P5} shows the temporal evolution of the absorption spectrum in the planet's rest frame during the transit for the Earth-sized case with a spot size of 1.25 times the planetary radius and a stellar rotation period of 5 days. We notice, around the spot-only line, that up to the mid-transit phase, the POLD is in absorption. After the mid-transit phase the absorption spectrum shows a mix of emission and absorption for the POLD until the end of the transit. During the spot crossing event (at a time $\approx$ 4.5 hours), the POLD is in emission, similar to the Jupiter-sized case. However, the spot crossing event is not dominant, most likely because the spot crossing event happens for a spot position closer to the stellar limb and is shorter than in the Jupiter-sized case. Most of the transit is in absorption, which results in an absorption signature in the mean absorption spectrum in Fig. \ref{fig:earth_abs_spec}. \\

\begin{figure}
    \centering
    \includegraphics[width=\linewidth]{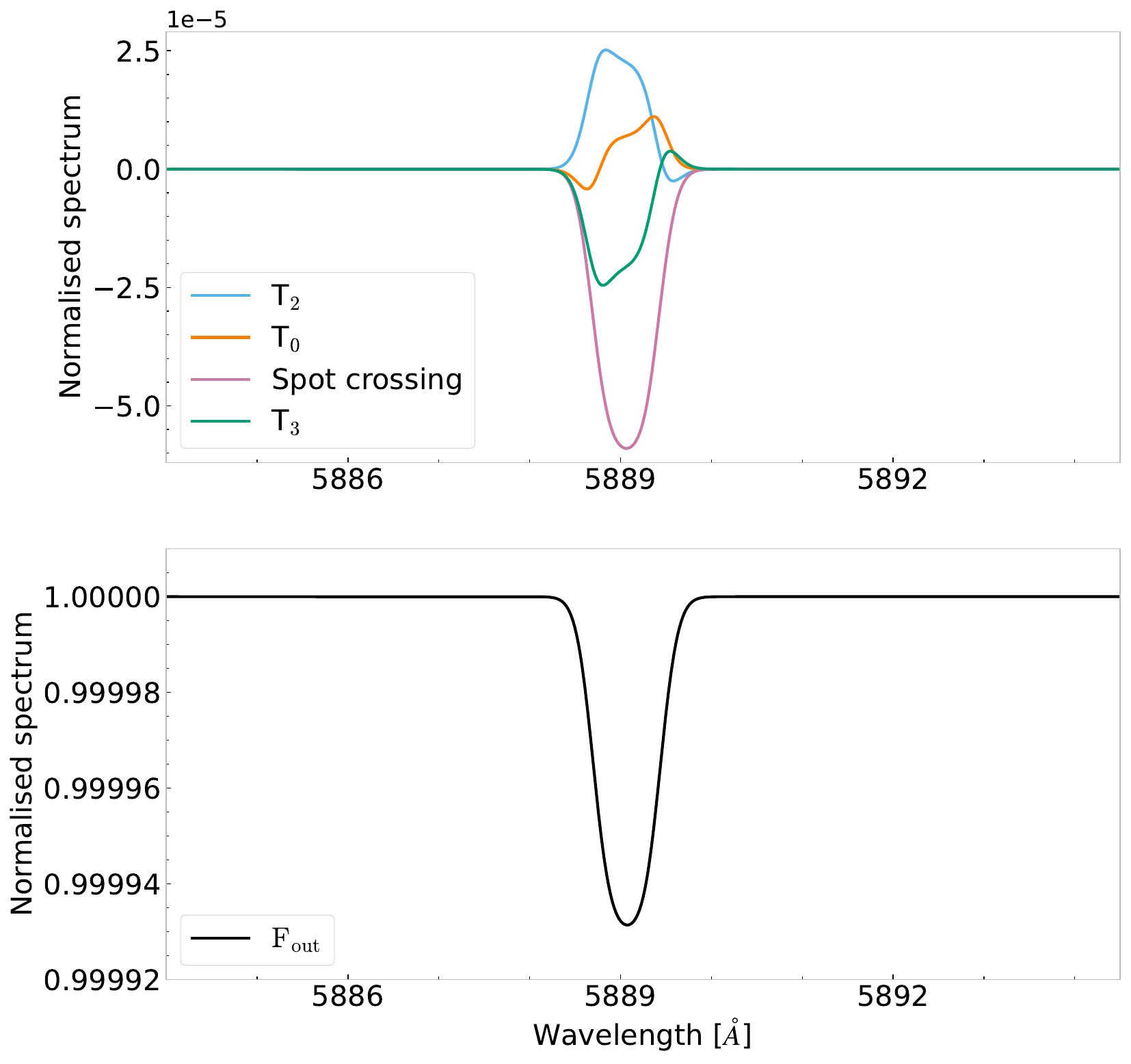} 
    \caption{Upper panel: Absorbed spectra at different symbolic time-steps as a function of wavelength for the case with $\rm s_p = 1.25$ and $\rm P_* = 5$ for the Earth-sized case. Lower panel: Normalised out-of-transit spectrum as a function of wavelength for the case with $\rm s_p = 1.25$ and $\rm P_* = 5$. 
    }
    \label{fig:earth_absorbed_spec}
\end{figure}

Figure \ref{fig:earth_absorbed_spec} shows the out-of-transit spectrum and absorbed spectra at different timesteps (same as for the Jupiter-sized case) as a function of wavelength for the Earth-sized case. For the spot crossing time step, the signal in $\rm F_{abs}$ is larger than for the time steps of non-spot-crossing because the local spectrum actually contains a contribution from the spot-only line.\\ 
For the non-spot-crossing time steps, the observed signal is only due to the discrepancy between $\rm F_{out}$ and $\rm F_{in}$, as explained in Sect. \ref{sec:jupiter}.
Moreover, in the Earth-sized case, given the much longer in-transit duration, the $\rm F_{out}$ was computed using very different spectra (before and after transit) due to the motion of the spot\footnote{In Appendix \ref{app:visualisation}, we show a visualisation of how much the spots have moved during the transit of the Jupiter-sized and Earth-sized cases for a stellar rotation period of 5 days} and is thus not representative of an instantaneous disc-integrated spectrum. Each in-transit disc-integrated spectrum $(\rm F_{in}$) is also different, not only due to the planet but also because of the spot motion. The interplay between these effects has a strong impact on the spectral distortions. 
As mentioned earlier, in the spectral region of the spot-only line, the disc-integrated out-of-transit spectrum contains only a tiny contribution from the spot due to the much smaller size of the spot than in the Jupiter-sized case. As a result, $\rm F_{out}$ appears almost as a constant spectrum, and the results presented in Fig. \ref{fig:earth_abs_spec_2D_sp125_P5} are thus almost directly tracing what is observed in $\rm F_{abs}$.\\

\begin{figure}
    \centering
    \includegraphics[width=\linewidth]{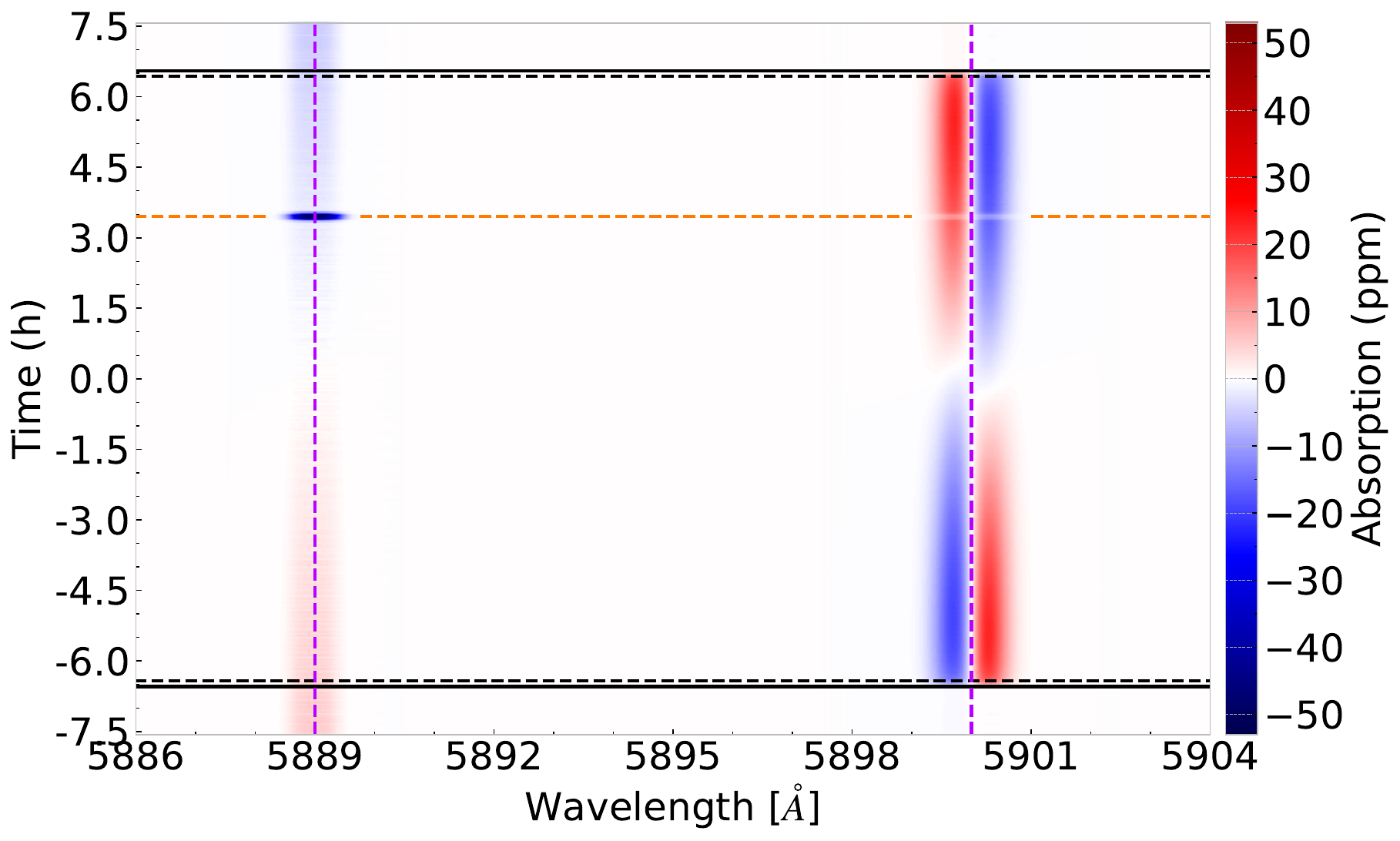}
    \caption{Absorption spectrum as a function of time and wavelength in the planet's rest frame for the case with $\rm s_p = 1.25$ and $\rm P_* = 25$ for an Earth-sized system. The horizontal plain and dashed black lines mark the contact times of the transit. The vertical dashed purple lines mark the reference wavelength of the stellar lines. The horizontal dashed orange line marks the spot crossing's time step.}
    \label{fig:earth_abs_spec_2D_sp125_P25}
\end{figure}

However, we note that for the case with the stellar rotation period of 25 days, the distortions are in emission, contrary to the cases with the two other periods. Figure \ref{fig:earth_abs_spec_2D_sp125_P25} shows the temporal evolution of the absorption spectrum during the transit for the Earth-sized case with a spot size of 1.25 times the planetary radius and a stellar period of 25 days. We see that the spot crossing happens earlier during the transit at around 3.5 hours. Therefore, the spot crossing event contributes more to the average absorption spectrum due to the more intense flux behind the planet. This is due to the closer proximity of the spot to the disc centre as the star rotates more slowly than in the case with P$_* = 5$ days. Furthermore, the amplitude of the distortions for the rest of the transit is less intense than for the spot crossing timestep, and there is more or less the same number of timesteps having absorption-like than emission-like distortions. The average absorption spectrum thus shows a distortion in emission for the spot-only line. 

\subsubsection{Master out-of-transit spectrum computed with only half of the baseline}
\label{sec:half_baseline}
\begin{figure}
    \centering
    \includegraphics[width=\linewidth]{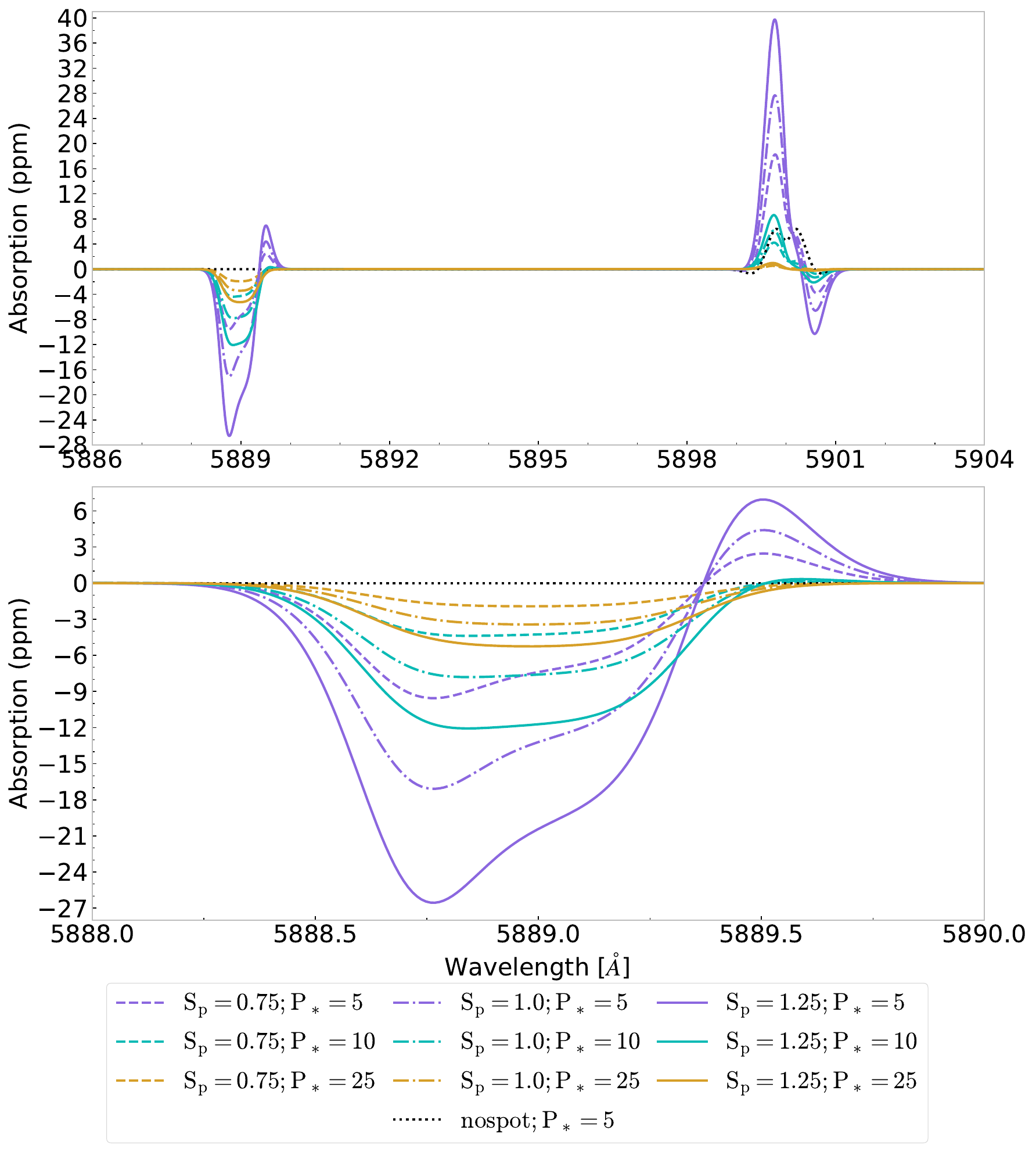}
    \caption{Upper panel: Absorption spectrum averaged between T$_2$ and T$_3$ in the planet's rest frame for the different cases for the Earth-sized system computed with the pre-transit-only $\rm F_{out}$. Lower panel: Zoom on the wavelength range around the spot spectral line.}
    \label{fig:pre}
\end{figure}

\begin{figure}
    \centering
    \includegraphics[width=\linewidth]{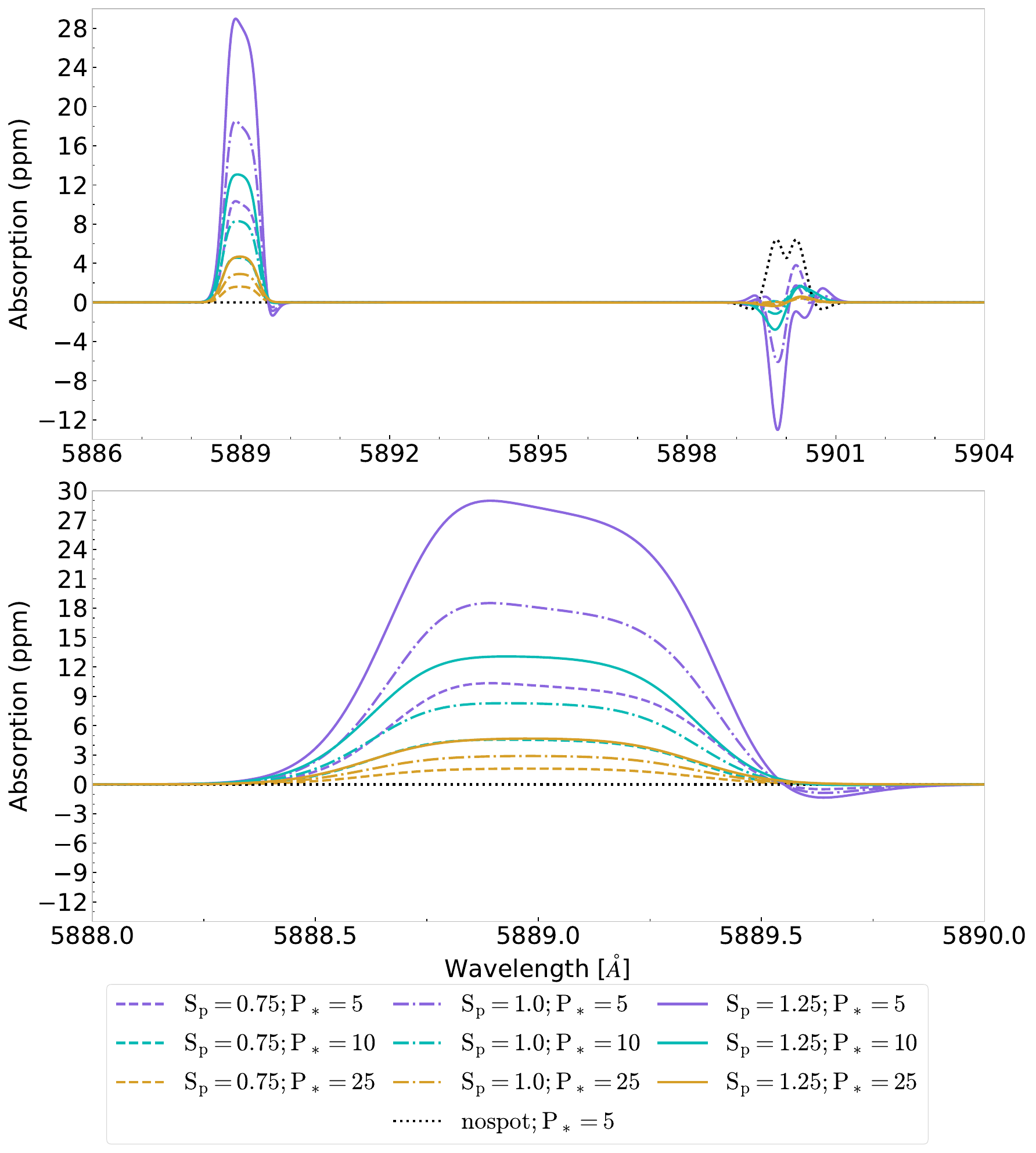}
    \caption{Upper panel: Absorption spectrum averaged between T$_2$ and T$_3$ in the planet's rest frame for the different cases for the Earth-sized system computed with the post-transit-only $\rm F_{out}$. Lower panel: Zoom on the wavelength range around the spot spectral line.}
    \label{fig:post}
\end{figure}

In the case of an Earth-sized planet on an Earth-like orbit, the transit duration is approximately 13 hours. Therefore, for most ground-based observations, one cannot observe the full transit plus baseline of a planet with the same orbital period as the Earth over a single night. Since the transit would only be partially observed, the baseline to compute the master out-of-transit spectrum would, at best, rely only on the 1-hour baseline before or after the transit. \\
Arguably, the spot-only line in the master out-of-transit spectrum should change if we compute it only with the pre- or post-transit exposures. This comes from the fact that for the full baseline case, the spot has moved more between the different exposures that compose the average spectrum. Therefore, with a full baseline, the disc-integrated spot-only line in $\rm F_{out}$ presents a more diluted (broader) profile. In the cases with only the before or after exposures, the disc-integrated line profile in the spot-only line should be less broad, more peaked, and shifted. With a less broad, more peaked, and shifted line profile in the $\rm F_{out}$, the POLDs should exhibit amplitudes and shape changes.\\
To illustrate this, we ran simulations for the Earth-sized case, defining the master out-of-transit spectrum using either only the before-transit or only the after-transit exposures. Figures \ref{fig:pre} and \ref{fig:post} show the mean absorption spectra obtained using the pre-transit-only $\rm F_{out}$ and the post-transit-only $\rm F_{out}$, respectively. 
We see that the amplitude and shape of the POLDs around the spot-only line have changed noticeably. The amplitude is larger in both cases than for the case using a master $\rm F_{out}$ defined with the before- and after-transit exposures (see Fig. \ref{fig:earth_abs_spec}). Part of the discrepancy between the two new cases is due to the fact that, although the spot-only line profile is different than the full baseline case, the change is not symmetric between the two cases. The spot-only line profile in the post-transit-only $\rm F_{out}$ is actually shallower than in the other case because the spot contribution is less important. Indeed, in this particular case with a positive spot longitude, the spot is closer to the limb during the post-transit exposures than the pre-transit ones. Due to limb darkening, the spot contribution to the disc-integrated spectrum is thus less important in that case. 

\subsection{Suppressing the motion of the spot} 
\begin{figure}
    \centering
    \includegraphics[width=1\linewidth]{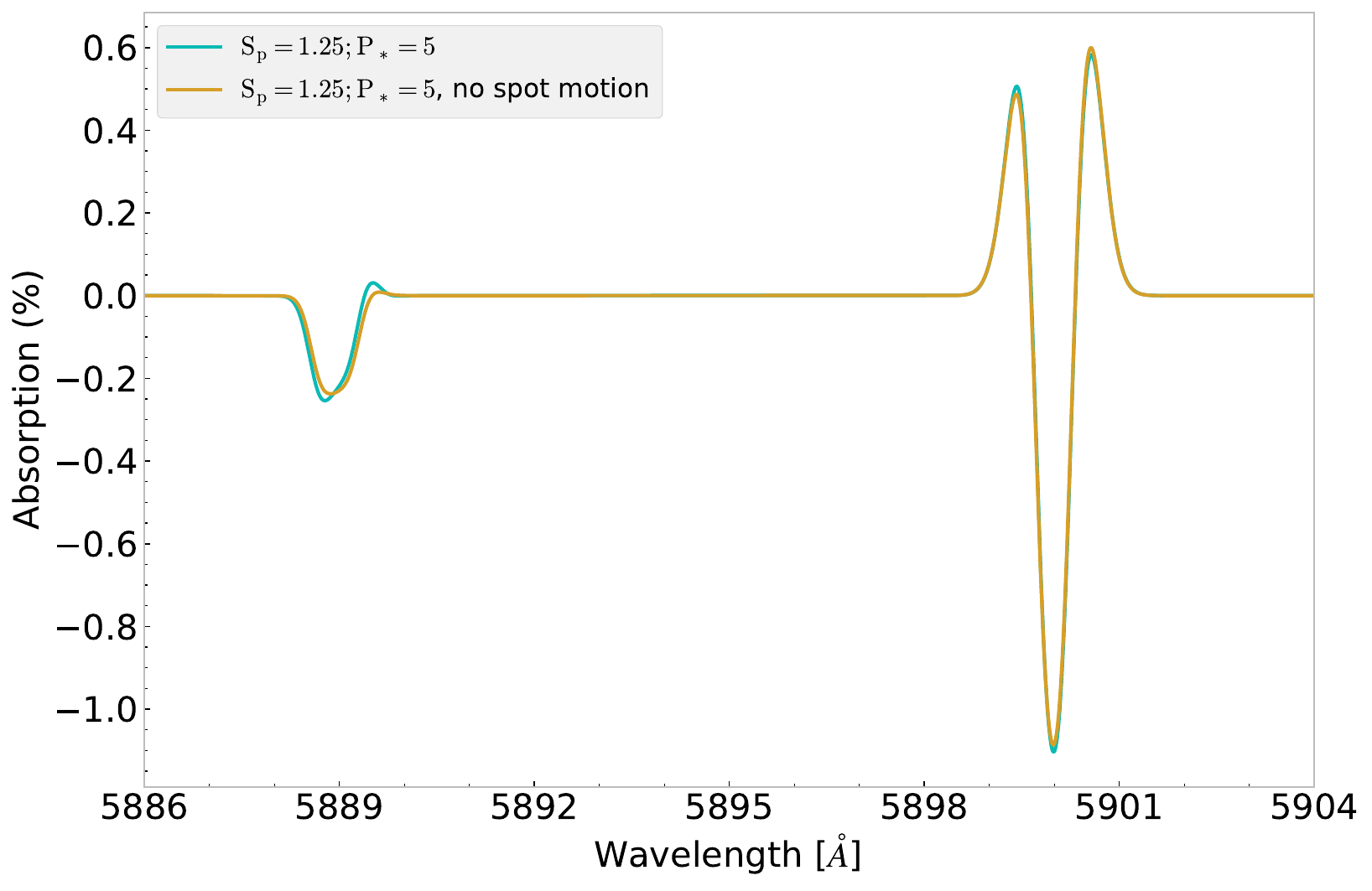}
    \caption{Absorption spectrum averaged between T$_2$ and T$_3$ in the planet's rest frame for the case with $\rm s_p = 1.25$, $\rm P_* = 5$ for a Jupiter-sized system. The gold-coloured curve shows the same simulation as the turquoise curve but with a static spot.}
    \label{fig:jupiter_abs_spec_mean_T23_nomotion_newLD}
\end{figure}

\begin{figure}
    \centering
    \includegraphics[width=1\linewidth]{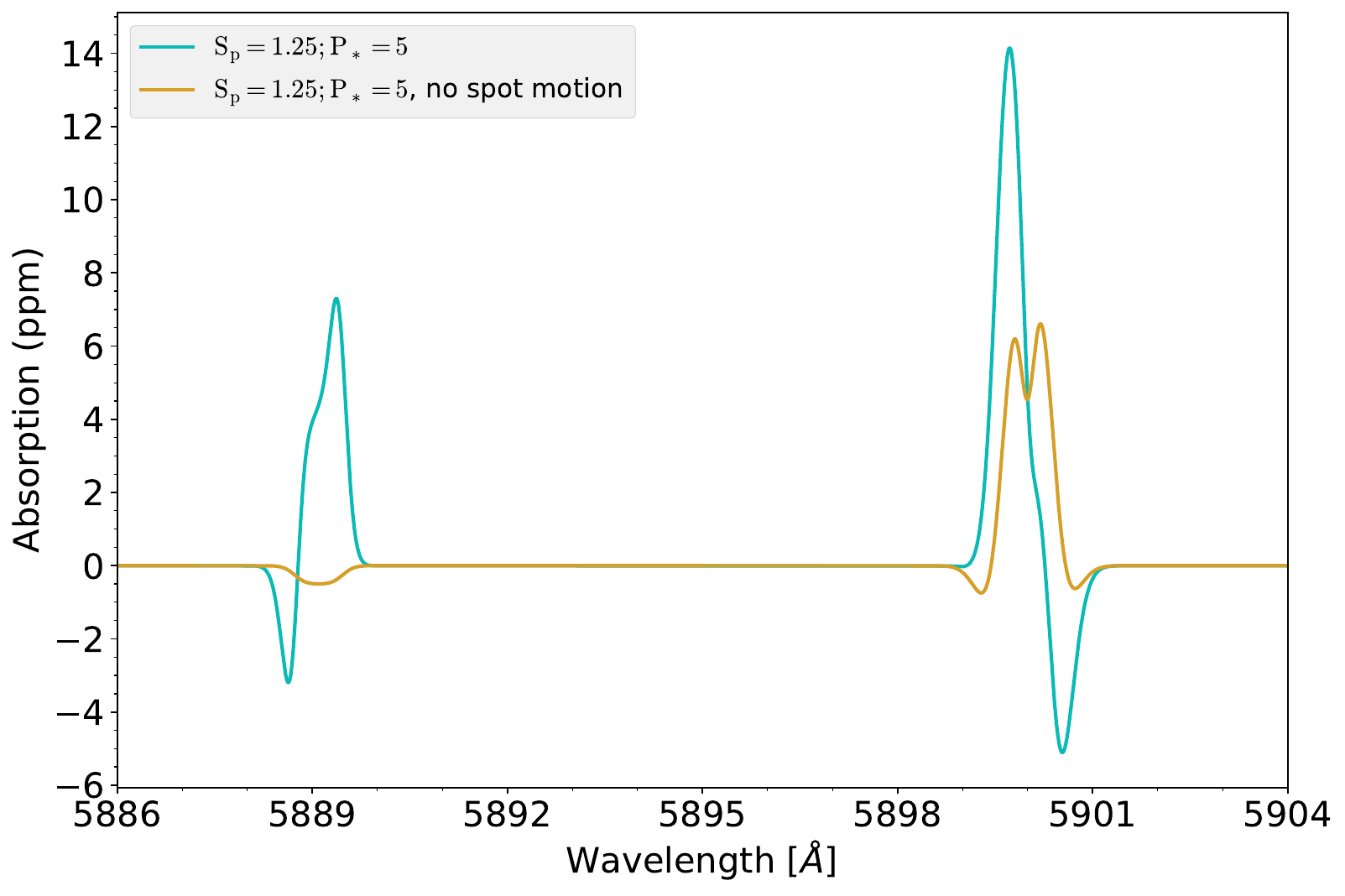}
    \caption{Absorption spectrum averaged between T$_2$ and T$_3$ in the planet's rest frame for the case with $\rm s_p = 1.25$, $\rm P_* = 5$ for an Earth-sized system. The gold-coloured curve shows the same simulation as the turquoise curve, but with a static spot.}
    \label{fig:earth_abs_spec_mean_T23_nomotion_newLD}
\end{figure}

To test the hypothesis that the spot motion is partly responsible for the distortions around the spot-only line, we simulated a new transit for both systems, with the spot radius of 1.25 times the planet's radius and the stellar rotation period of 5 days, our most extreme case, but this time disabling the motion of the spot in the simulation. This means that the spot remains at the initial position of longitude = 30$^\circ$ and latitude = 0$^\circ$. So there was a spot crossing event, but the spot remained at the same position during the whole simulation. Figures \ref{fig:jupiter_abs_spec_mean_T23_nomotion_newLD} and \ref{fig:earth_abs_spec_mean_T23_nomotion_newLD} show the mean absorption spectra between T$_2$ and T$_3$ for the Jupiter-sized and Earth-sized case, respectively, with and without the spot motion included in the simulation. \\
For the Jupiter-sized case, the small difference between the simulation with and without spot motion is explained by the fact that even with spot motion, the spot does not move significantly during the transit due to the short orbital period of the planet compared to the rotation period of the star (see also Fig. \ref{fig:spot_motion_earth}). The amplitude of the distortion remains similar to the case with spot motion, as the spot size is quite large with respect to the star's radius, and thus still induces significant distortion in the line profiles. \\
The difference between the simulations with and without the spot motion is much more striking for the Earth-sized case than for the Jupiter-sized case.
For the Earth-sized case, when the spot motion is inhibited, the signal around the spot-only line is drastically reduced in amplitude, peaking at $\sim$ -0.5 ppm. For the non spot-crossing exposures, the spot-only line is the same in $\rm F_{out}$ and $ \rm F_{in}$, apart from a difference in the flux level, so there is barely any distortion created for these exposures. The dominating distortion comes from the spot crossing exposures, contrary to the cases with spot motion.\\
These results indicate that one of the driving mechanisms for the large distortions around the spot-only line (especially for the long orbital period case) is indeed the motion of the spot, and the discrepancies that it creates between all the different disc-integrated spectra. This result also emphasises the importance of accounting for the motion of the spot to accurately simulate absorption spectra with active regions. Indeed, the sole presence of the spot would not be sufficient for reproducing the full spectral distortion of the absorption spectrum.\\
Another aspect to consider with real observations would be the smearing effect of the spot's shape in each in-transit exposure due to time averaging, increasing the effective size of the spot during an exposure. This effect was not considered here, but could become significant in cases with low stellar rotation periods. 

\section{Non-proportional spot sizes}
\label{sec:spotsize}
So far, we have considered spot sizes proportional to the radius of the simulated planet, but there was no physical motivation behind this choice. So we decided to run simulations for the close-in Jupiter-sized planet with spots whose radii were scaled to those adopted for the Earth-sized simulations, to see what the impact is on the absorption spectrum.\\
For the Earth-sized case, it is less straightforward because using spots of the size of Jupiter induces a distortion in the radial velocity (RV) curve of the system that is greater than the distortion induced by the Keplerian motion of the planet or even greater than the Rossiter-McLaughlin (RM) effect \citep{rossiter1924,mclaughlin1924}, such as in Fig. 3 of \citet{palle2020}. Therefore, when looking at the RV curve, a large non-linear distortion becomes noticeable, and it becomes obvious that a spot is present on the visible stellar surface \citep[e.g.][]{palle2020}. \\
In practice, fitting a linear function to the out-of-transit RV points (as we have done so far in this study) would not be sufficient and a more detailed treatment of this activity-induced distortion would need to be performed \citep{palle2020}. As such, we decided not to explore such cases, as the methodology we would have to use is outside the scope of this study and would differ too much from the methodology we used for the rest of the cases\footnote{Additional distortions are expected in the stellar lines because the observations are not averaged in the true stellar rest frame; locally, the lines of the quiet stellar regions appear slightly misaligned in the integrated spectra.}. Instead, we decided to run new simulations for the Earth-sized case with increasing spot sizes until the distortion in the RV curve caused by the spot started to noticeably modulate the planetary signal (see Sect. \ref{sec:earth_largespot}). 

\subsection{Close-in Jupiter-sized planet and Earth-sized spots}
\begin{figure}
    \centering
    \includegraphics[width=1\linewidth]{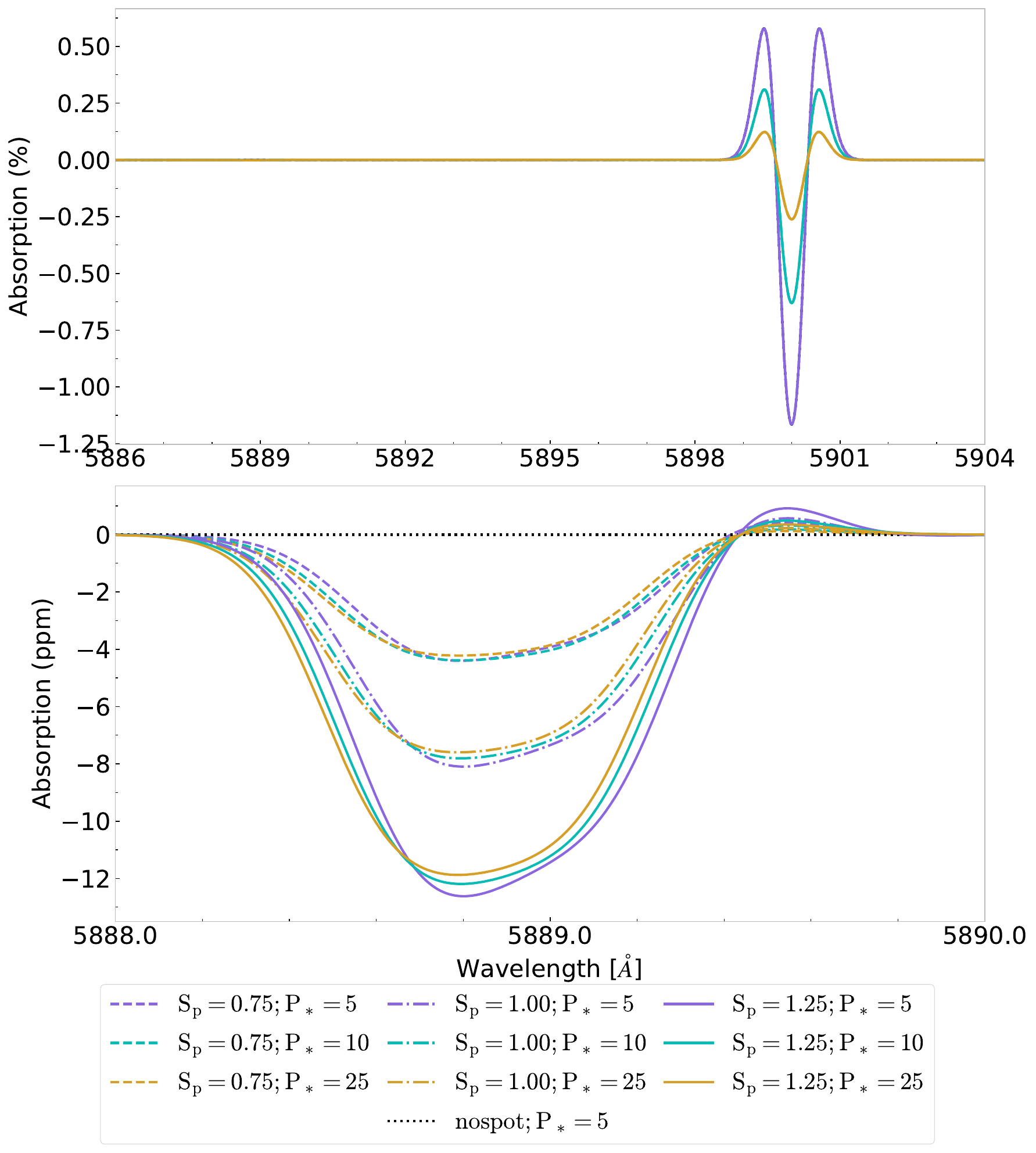}
    \caption{Upper panel: Absorption spectrum averaged between T$_2$ and T$_3$ in the planet's rest frame for the different cases for the Jupiter-sized system and Earth-sized spots. Lower panel: Zoom on the wavelength range around the spot spectral line. The spot sizes, $\rm S_p$ are given in units of the Earth's radius.}
    \label{fig:jupiter_abs_spec_invspotsize}
\end{figure}

Figure \ref{fig:jupiter_abs_spec_invspotsize} shows the results of the simulations for the close-in Jupiter-sized planet and spot sizes proportional to the Earth's radius. Quite as expected, the POLD induced by the spot-only line is much smaller than the one induced by the photospheric line, to the point that it is negligible when directly comparing them. The amplitude of the POLD now ranges from a few to a dozen parts per million, meaning it has decreased by a factor of $\sim$ 200 compared to the cases with spot sizes comparable to Jupiter's size. However, despite the change in amplitude, the behaviour of the POLD with respect to the spot size and stellar rotation period remains the same; that is, dominated by the size of the spot, increasing with spot size, and increasing with decreasing stellar rotation period.

\subsection{Distant Earth-sized planet and larger spots}
\label{sec:earth_largespot}
\begin{figure}
    \centering
    \includegraphics[width=\linewidth]{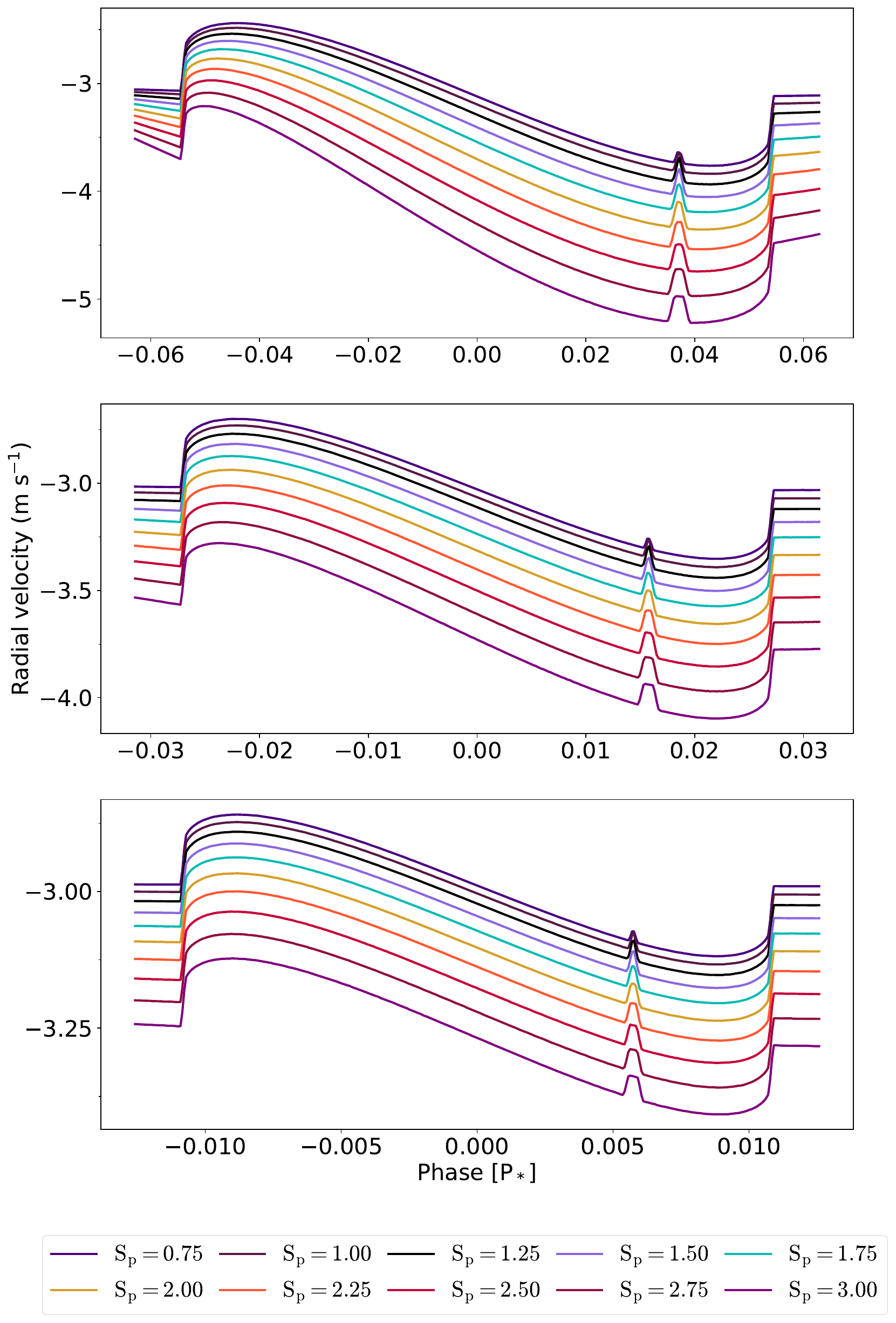}
    \caption{Radial velocity as a function of orbital phase for the different spot sizes and stellar rotation periods for the Earth-sized system. The curve contains the motion of the star around the barycentre, the spot induced RV distortion, and the RM effect. Upper panel: P$_* = 5$ days. Middle panel: P$_* = 10$ days. Lower panel: P$_* = 25$ days. The spot sizes are given in units of the Earth's radius.}
    \label{fig:rv}
\end{figure}

\begin{figure}
    \centering
    \includegraphics[width=\linewidth]{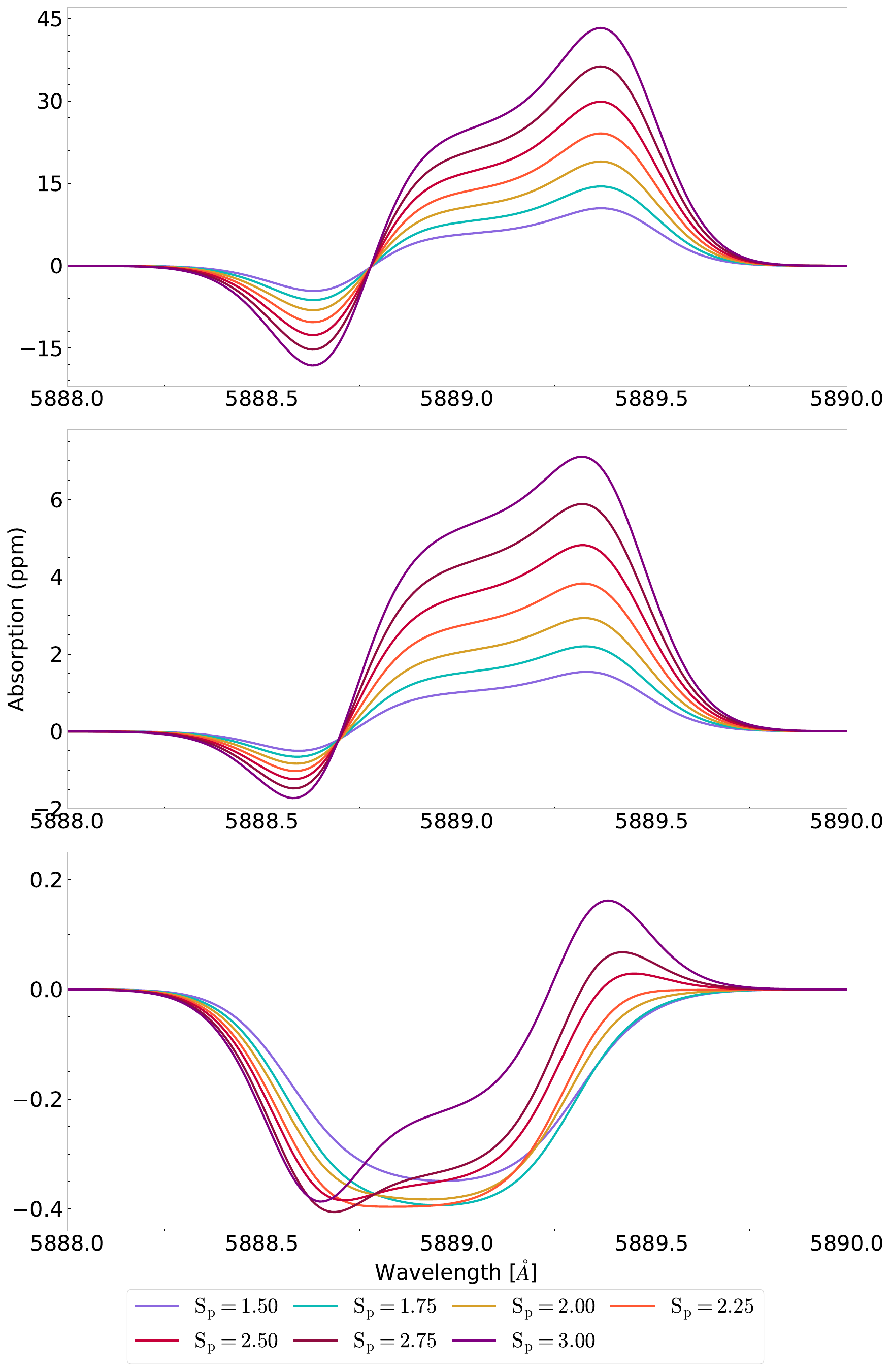}
    \caption{Absorption spectrum averaged between T$_2$ and T$_3$ in the planet's rest frame for the different cases for the Earth-sized system and larger spots than the initial simulations. Upper panel: P$_* = 5$ days. Middle panel: P$_* = 10$ days. Lower panel: P$_* = 25$ days. The spot sizes, $\rm S_p$ are given in units of the Earth's radius.}
    \label{fig:earth_abs_spec_largerspotsize}
\end{figure}

Figure \ref{fig:rv} shows the radial velocities in the stellar rest frame as a function of orbital phase of the planet for the different sizes and stellar rotation periods for the case of an Earth-sized planet. The typical RM effect is clearly visible, and the spot crossing event is also noticeable as a small peak in the RV curve.\\
As the size of the spot increases and the stellar rotation period decreases, we notice that the RV curve starts to be modulated. The case with spot size $\rm S_p = 3.0$ and P$_* = 5$ days starts to resemble the behaviour in Fig. 3 of \citet{palle2020}. We therefore decided not to go beyond spot sizes greater than 3 times the planetary radius. Figure \ref{fig:earth_abs_spec_largerspotsize} shows the results of the simulations for the long orbital period Earth-sized planet and larger spot sizes proportional to the Earth's radius, ranging from $\rm S_p = 1.5$ to 3 by steps of 0.25. As already discussed in Sect. \ref{sec:earth}, increasing the size of the spot leads to an increase in the amplitude of the POLDs around the spot-only line in the absorption spectrum. We also note that, as the spot size increases, for the case with P$_* = 25$ days, the behaviour of the POLD starts to approach that of the other stellar rotation periods. \\
A plausible explanation is that, during a transit, for a given P$_*$, a larger spot covers spatially a broader range of blue- and red-shifted stellar surfaces than a smaller spot. This effect on the spot-only line is analogous to that produced by a shorter P$_*$, since, for a fixed spot size, a faster rotation also increases the range of Doppler-shifted regions covered by the spot. In both cases, the consequence is a similar broadening effect of the spot-only line in the out-of-transit disc-integrated spectrum.
 
\section{Conclusions}
\label{sec:conclusion} 
Due to their lower temperature compared to the photosphere, spots can display absorption lines in their spectrum that are absent from the photosphere's spectrum. We investigated the impact of such spot-only lines on the high-resolution absorption spectrum of a transiting planet. For this purpose, we used the SOAPv4 code to simulate the transit of two different planets in front of a star with a dark spot: a distant Earth-sized planet and a close-in Jupiter-sized planet. To study the variation in the spot-induced distortion, we explored different configurations by varying the spot size and the rotation period of the star. We found that under certain conditions, the distortions induced by the spot-only line can significantly affect the absorption spectrum of the planets, in the case of crossing and non-spot-crossing events. \\
The exact amplitudes of these spot-induced distortions should of course be interpreted with caution. Given the ideal and streamlined setup that we used, it is probable that their exact values differ from the results of the present simulations.\\

Nevertheless, under the assumptions that we made, for an Earth-sized planet on a long-period orbit around a star with a rotation period similar to the Sun's, the distortion introduced in the absorption spectrum by the spot-only line is in emission and on the order of a few tenths of a part per million, which could dampen actual absorption signatures from the atmosphere. For shorter rotation periods than the Sun's, we showed that the distortion signature introduced by the spot is in absorption and of an order of magnitude comparable to the expected amplitude of an Earth-Sun system atmosphere signature \citep[i.e. a few parts per million,][]{ehrenreich2006,Ardaseva2017,Kreidberg2017}. Such distortions could be misinterpreted as the signature of an element from the planetary atmosphere. For the close-in Jupiter-sized planet, the distortion introduced in the absorption spectrum by the spot is in emission and of a few tenths of a percent, and could thus possibly dampen a planetary atmosphere absorption signature.\\

Additionally, the spurious signatures originating from the spot-only line can affect nearby planetary absorption lines in the absorption spectrum. This can happen when a planetary atmosphere absorption feature is shifted due to, for example, day-to-night side winds in the planetary atmosphere.

The main driver for the origin of this feature is not solely the spot crossing event in itself, but the fact that the master out-of-transit spectrum is a poor estimate of the local spectra absorbed by the planet, which is one of the major limitations of transit spectroscopy. In this particular study, this was due to the spot moving significantly during the out- and in-transit phases. The computation of a master out-of-transit spectrum using before and after exposures is thus not an optimal choice. Moreover, each in-transit spectrum is also changing exposure by exposure due to the motion of the spot, adding to the discrepancy between the out-of- and in-transit spectra used to compute absorption spectra. This outcome is amplified for long-period planets, as the spot has more time to move over the visible surface of the star. There is thus a need for detailed modelling of the stellar surface in and out of transit to correctly account for the spot-induced distortions in high-resolution absorption spectra. Alternatively, one could model, for each exposure, the in-transit disc-integrated spectra without planetary absorption, to use as an alternative to $\rm F_{out}$. Moreover, when such activity features are detected in transit observations, one way to disentangle the true planetary signature from the activity-induced POLDs would be to gather additional observations for different nights in the hope that the activity region is not present during these other transits.\\

This study was conducted using a single line with a standard shape and no CLV in the shape of line profiles; therefore, depending on the actual studied line profiles, the exact amplitude and shape of the resulting distortions should vary. We also assumed a spot latitude of 0$^\circ$, more simulations with non-zero latitude and different geometries should thus be performed in the future to understand how the distortions evolve. Including the Zeeman effect would also be interesting, as spots have strong magnetic fields, and as this effect can really alter the actual line profile, which ultimately would lead to a different shape in the distortions in the absorption spectrum. 

Finally, with the upcoming Paranal solar ESPRESSO Telescope (PoET) \citep{santos2025}, we shall conduct spatially resolved observations of sunspots and faculae and inject them into SOAPv4 to run simulations with realistic spot spectra to verify if our results hold. We shall also compare local spectra from both the photosphere and active regions to identify lines that would be suited for our science case; that is, ones that are strong in the active region's spectrum and faint or absent from the photosphere's spectrum. With these real spectra, we shall also naturally capture the effects of the magnetic field on the spot lines, such as the Zeeman splitting of lines, which could possibly have a strong impact on the resulting distortions of the absorption spectrum of the transiting planet.

\begin{acknowledgements}
This work was funded by the European Union (ERC, FIERCE, 101052347). Views and opinions expressed are however those of the author(s) only and do not necessarily reflect those of the European Union or the European Research Council. Neither the European Union nor the granting authority can be held responsible for them.

This work was supported by Fundação para a Ciência e a Tecnologia (FCT) through national funds under the research grant UID/04434/2025 (DOI 10.54499/UID/04434/2025). 

\end{acknowledgements}
\bibliographystyle{aa}
\bibliography{bibliography.bib}
\begin{appendix}
\section{Visualisation of the spot's motion during the transits}
\label{app:visualisation}
\begin{figure}[!htbp]
    \centering
    \includegraphics[width=1\linewidth]{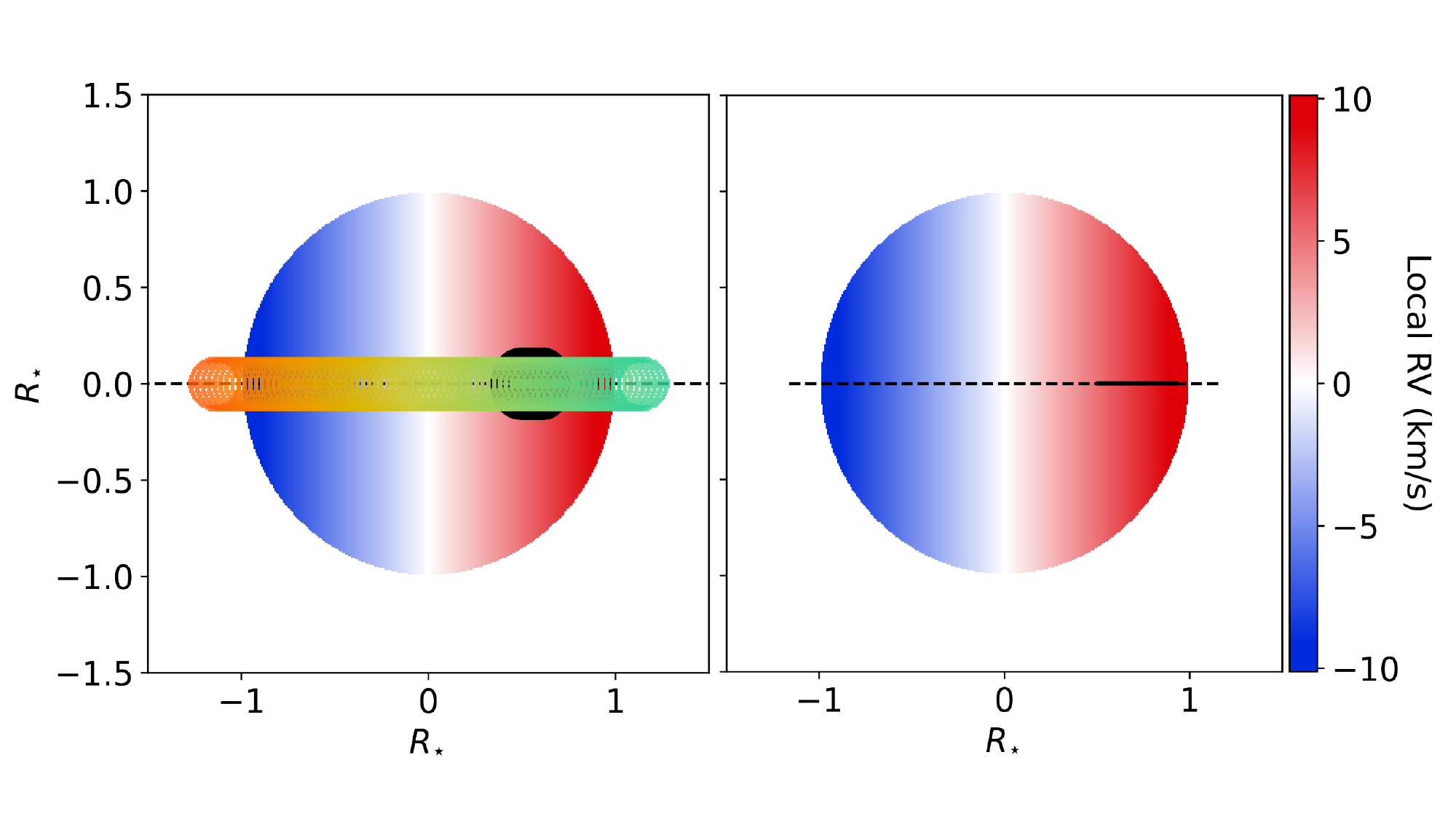}
    \caption{Visual representation of the motion of the spot on the stellar surface during the transit for P$_* = 5$ days and $\rm s_p=1.25$ times the planetary radius. Left panel: Close-in Jupiter-sized case. Right panel: Long orbital period, Earth-sized case. The spot is represented by the black discs that overlap. The coloured circles, although not visible for the Earth-sized case due to being too small, are the positions of the planet throughout the transit.}
    \label{fig:spot_motion_earth}
\end{figure}
\FloatBarrier
\section{Spot starting position variation}

\label{app:position}
\begin{figure}[!htbp]
    \centering
    \includegraphics[width=\linewidth]{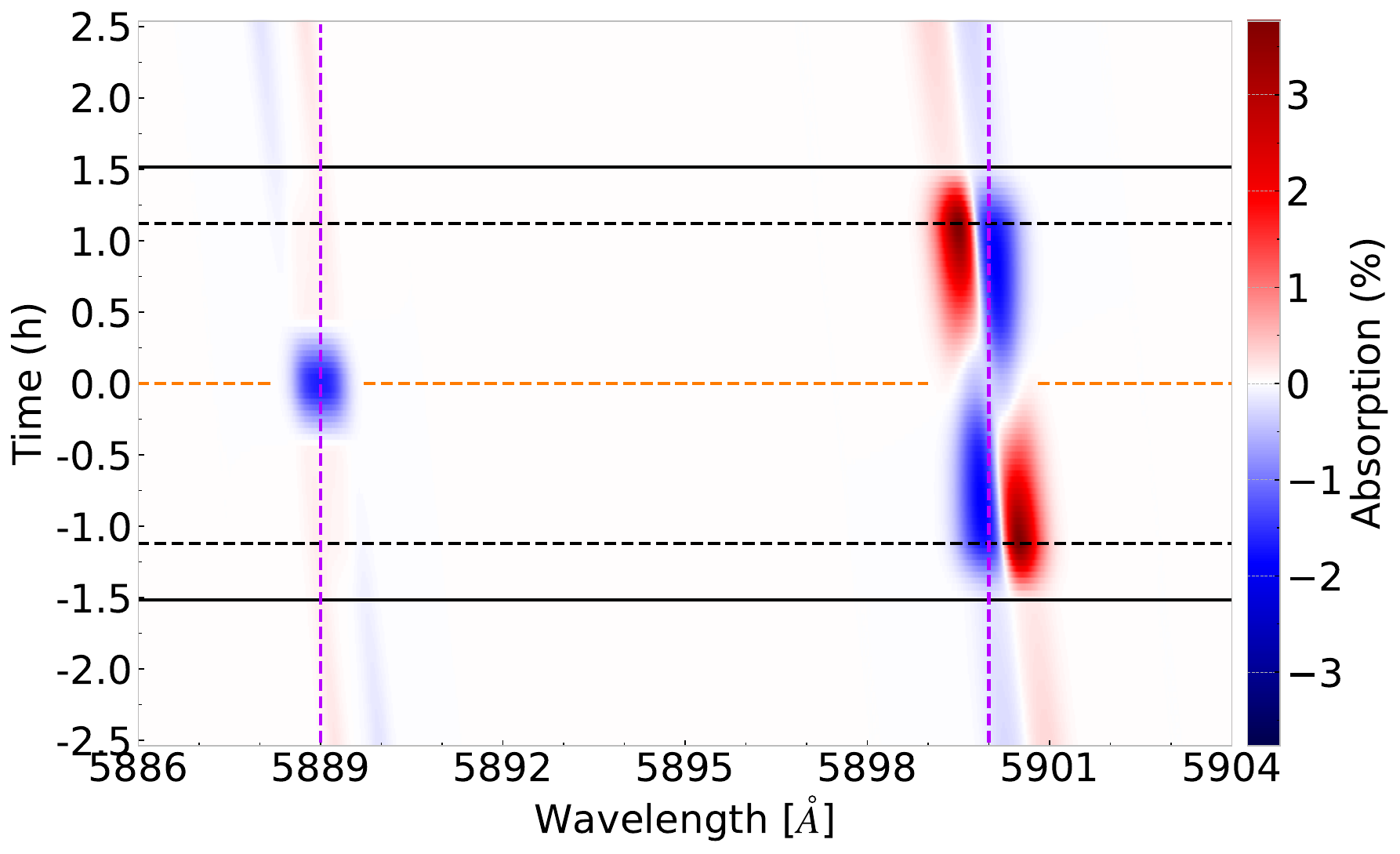}
    \caption{Absorption spectrum as a function of time and wavelength in the planet's rest frame for the case with $\rm s_p = 1.25$, a spot longitude of $0^\circ$ at T$_0$ and $\rm P_* = 5$ for a Jupiter-sized system. The horizontal plain and dashed black lines mark the contact times of the transit. The vertical dashed purple lines mark the reference wavelength of the stellar lines. The horizontal dashed orange line marks the spot crossing's time step.}
    \label{fig:jupiter_abs_spec_2D_sp125_l0_P5}
\end{figure}

\begin{figure}[!htbp]
    \centering
    \includegraphics[width=\linewidth]{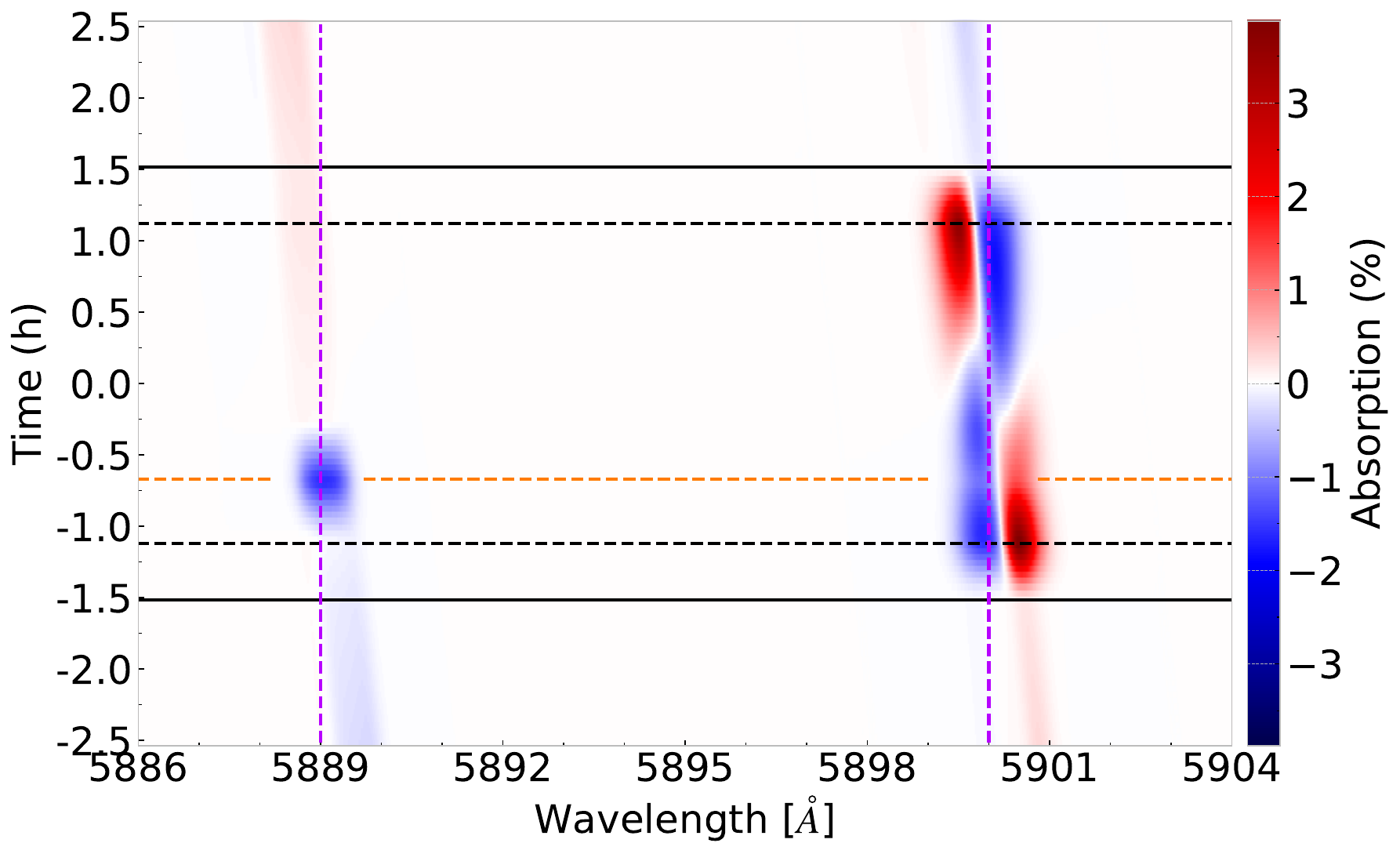}
    \caption{Absorption spectrum as a function of time and wavelength in the planet's rest frame for the case with $\rm s_p = 1.25$, a spot longitude of $-30^\circ$ at T$_0$ and $\rm P_* = 5$ for a Jupiter-sized system. The horizontal plain and dashed black lines mark the contact times of the transit. The vertical dashed purple lines mark the reference wavelength of the stellar lines. The horizontal dashed orange line marks the spot crossing's time step.}
    \label{fig:jupiter_abs_spec_2D_sp125_l_30_P5}
\end{figure}

\begin{figure}[!htbp]
    \centering
    \includegraphics[width=\linewidth]{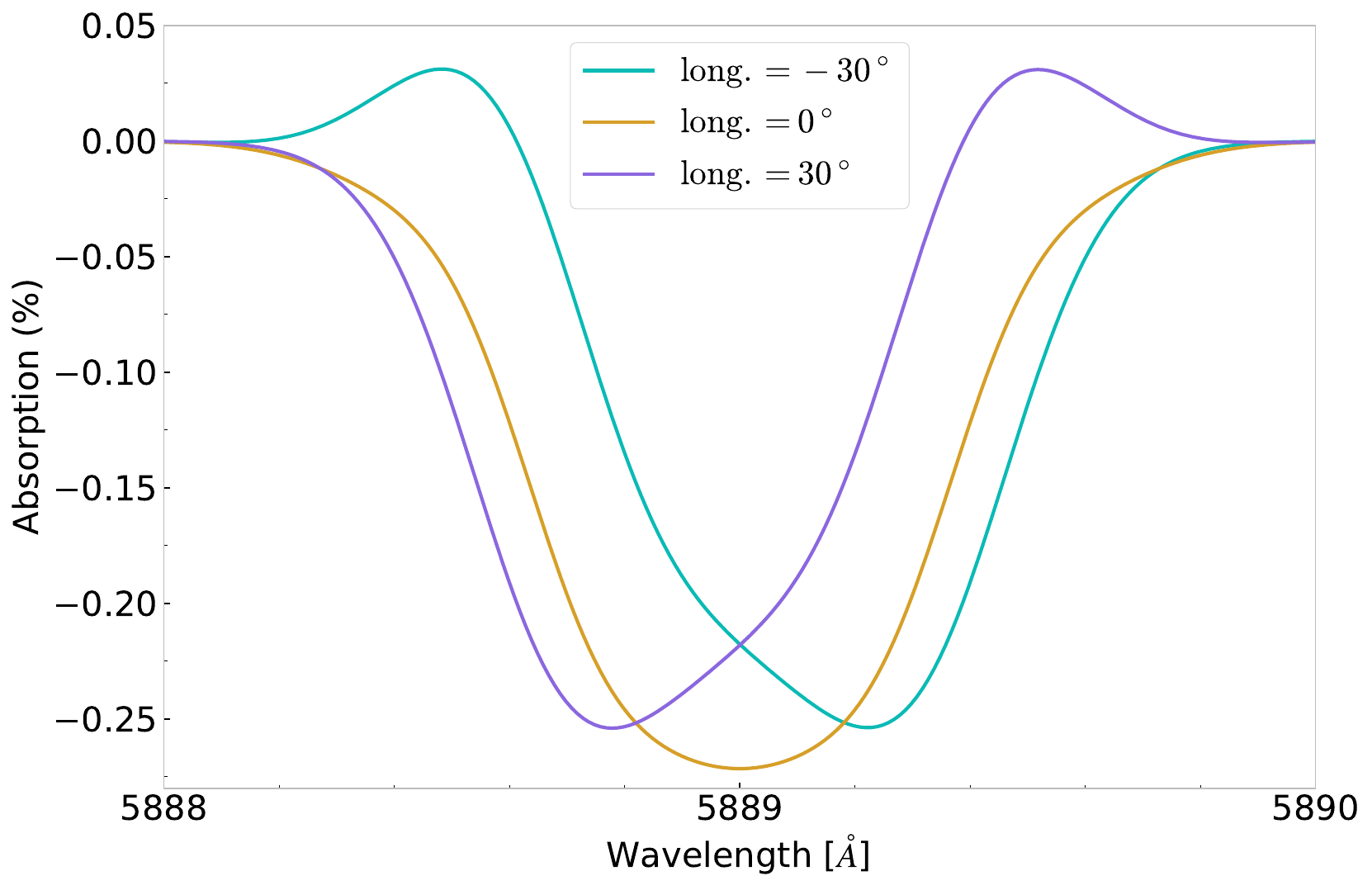}
    \caption{Absorption spectrum averaged between T$_2$ and T$_3$ in the planet's rest frame for the case with $\rm s_p = 1.25$, $\rm P_* = 5$ and different spot longitudes at T$_0$ for a Jupiter-sized system.}
    \label{fig:jupiter_abs_spec_mean_T23_sp125_l_P5}
\end{figure}

\begin{figure}[!htbp]
    \centering
    \includegraphics[width=\linewidth]{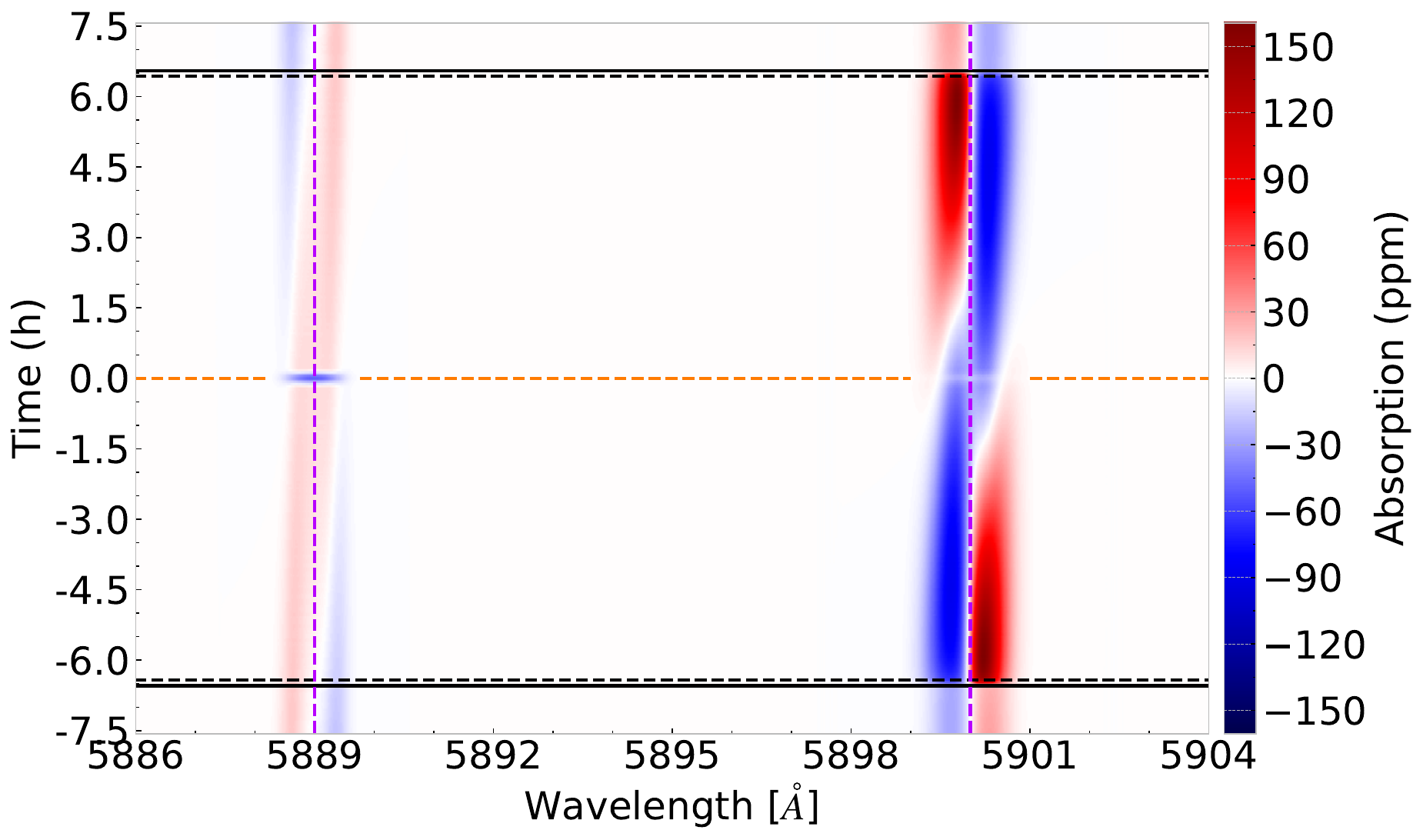}
    \caption{Absorption spectrum as a function of time and wavelength in the planet's rest frame for the case with $\rm s_p = 1.25$, a spot longitude of $0^\circ$ at T$_0$ and $\rm P_* = 5$ for an Earth-sized system. The horizontal plain and dashed black lines mark the contact times of the transit. The vertical dashed purple lines mark the reference wavelength of the stellar lines. The horizontal dashed orange line marks the spot crossing's time step.}
    \label{fig:Earth_abs_spec_2D_sp125_l0_P5}
\end{figure}

\begin{figure}[!htbp]
    \centering
    \includegraphics[width=\linewidth]{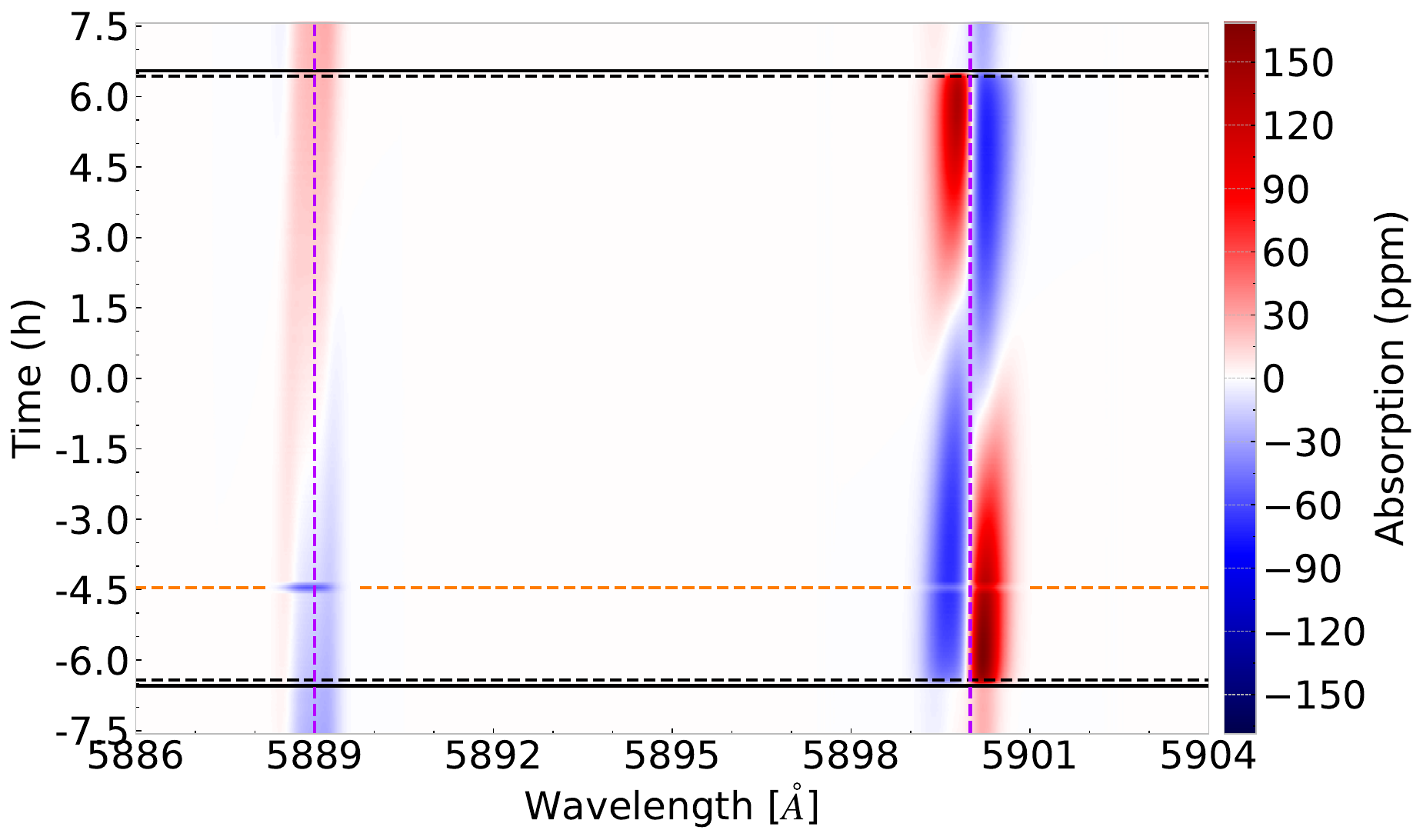}
    \caption{Absorption spectrum as a function of time and wavelength in the planet's rest frame for the case with $\rm s_p = 1.25$, a spot longitude of $-30^\circ$ at T$_0$ and $\rm P_* = 5$ for an Earth-sized system. The horizontal plain and dashed black lines mark the contact times of the transit. The vertical dashed purple lines mark the reference wavelength of the stellar lines. The horizontal dashed orange line marks the spot crossing's time step.}
    \label{fig:earth_abs_spec_2D_sp125_l_30_P5}
\end{figure}

\begin{figure}[!htbp]
    \centering
    \includegraphics[width=\linewidth]{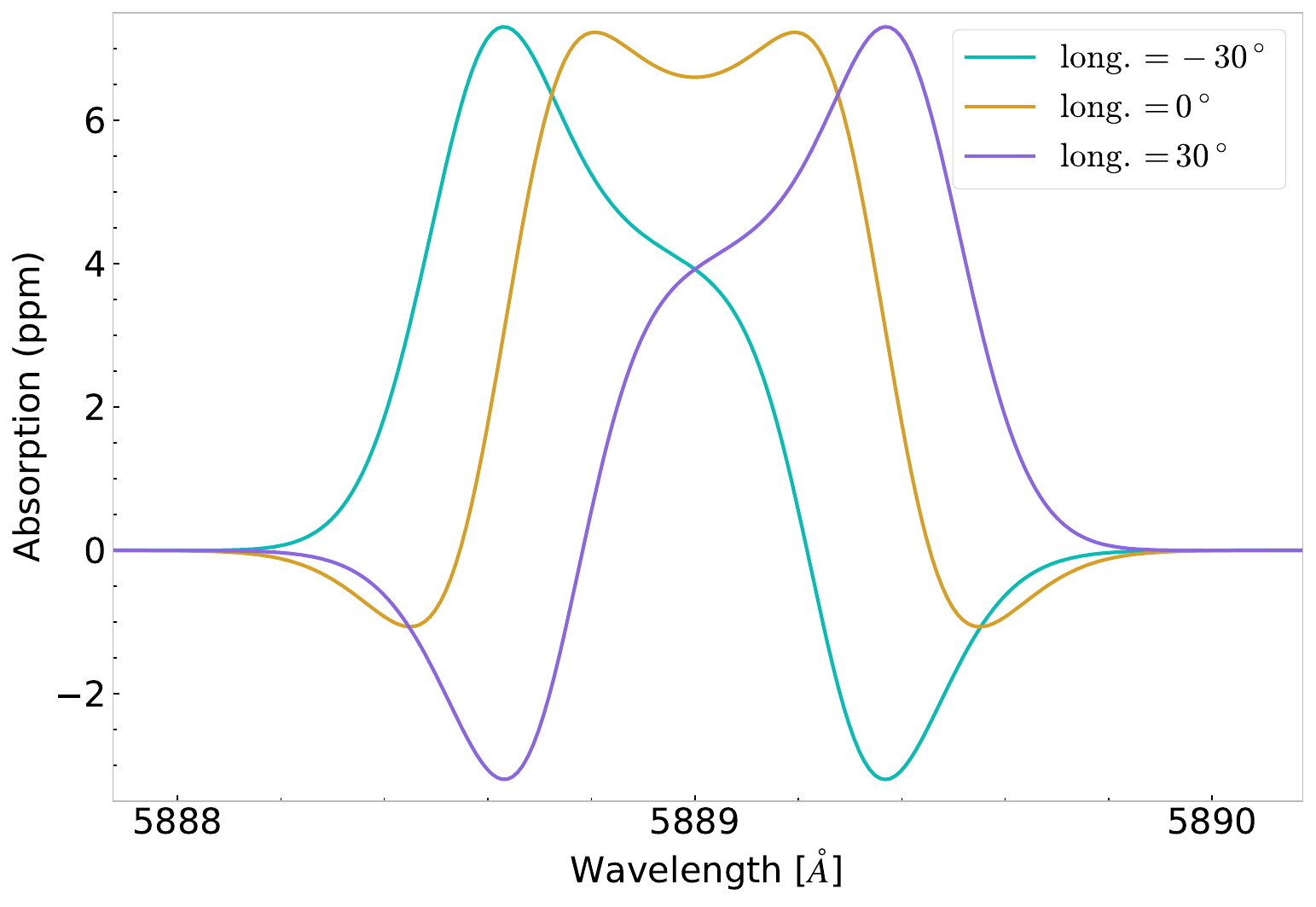}
    \caption{Absorption spectrum averaged between T$_2$ and T$_3$ in the planet's rest frame for the case with $\rm s_p = 1.25$, $\rm P_* = 5$ and different spot longitudes at T$_0$ for an Earth-sized system.}
    \label{fig:earth_abs_spec_mean_T23_sp125_l_P5}
\end{figure}

In this section, we changed the starting position of the spot in the simulations to see the impact on the resulting distortions in the absorption spectrum. The new positions were longitude = -30$^\circ$ and 0$^\circ$ at the middle of the transit; the latitude was set to 0$^\circ$. The rest of the parameters were set as in our most extreme case, i.e. a spot size of 1.25 times the planetary radius and a stellar rotation period of 5 days. Figures \ref{fig:jupiter_abs_spec_2D_sp125_l0_P5} and \ref{fig:jupiter_abs_spec_2D_sp125_l_30_P5} show the absorption spectra as a function of time for the Jupiter-sized case for the spot longitude = 0$^\circ$ and -30$^\circ$, respectively. Figure \ref{fig:jupiter_abs_spec_mean_T23_sp125_l_P5} shows the mean absorption spectrum between T$_2$ and T$_3$ for the Jupiter-sized case and the three different spot positions. Figures \ref{fig:Earth_abs_spec_2D_sp125_l0_P5} and \ref{fig:earth_abs_spec_2D_sp125_l_30_P5} show the absorption spectra as a function of time for the Earth-sized case for the spot longitude = 0$^\circ$ and -30$^\circ$, respectively. Figure \ref{fig:earth_abs_spec_mean_T23_sp125_l_P5} shows the mean absorption spectrum between T$_2$ and T$_3$ for the Earth-sized case and the three different spot positions. We notice, in every case, that the sign of the induced distortion by the spot-only line remains the same as in the initial case with longitude = 30$^\circ$. The amplitudes remain of the same order of magnitude. The shapes slightly vary.

\FloatBarrier
\section{Spot line depth variation}
\begin{figure}[!htbp] 
    \includegraphics[width=\linewidth]{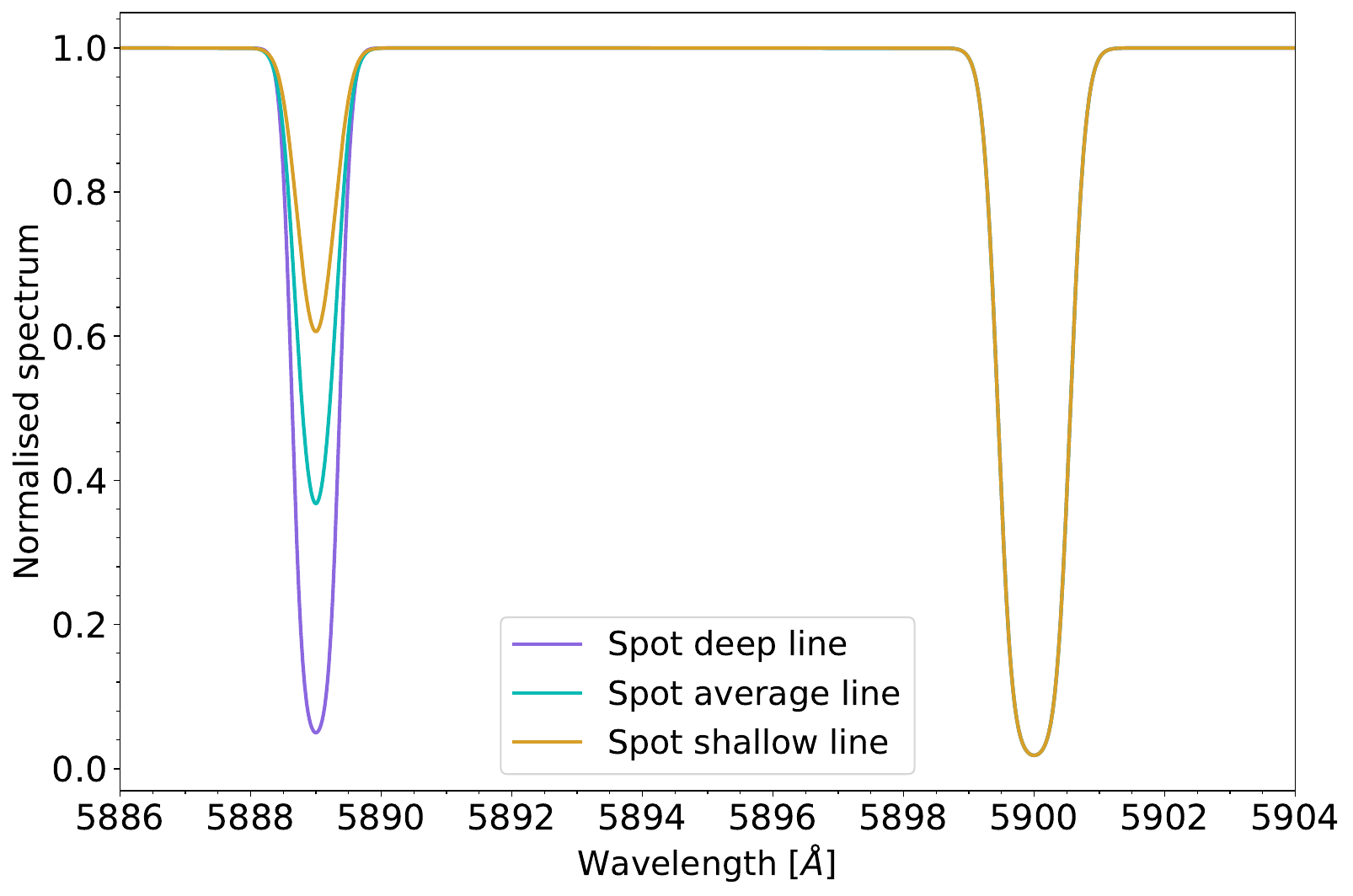}
    \caption{Input spectra used in SOAPv4 for the spot with different line depths.}
    \label{fig:spectrum}
\end{figure}

\begin{figure}[!htbp]
    \centering
    \includegraphics[width=\linewidth]{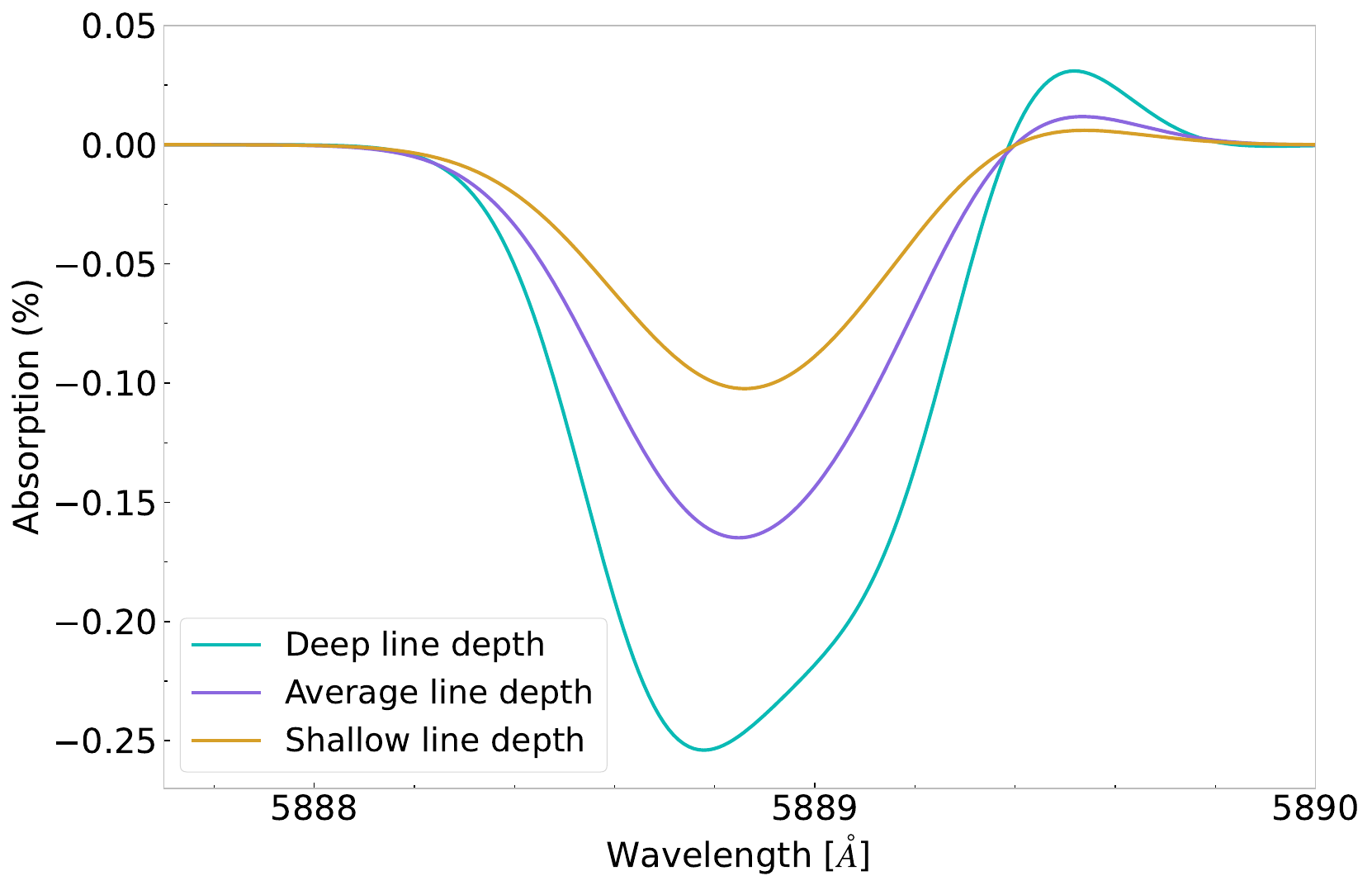}
    \caption{Absorption spectrum averaged between T$_2$ and T$_3$ in the planet's rest frame for the case with $\rm s_p = 1.25$, $\rm P_* = 5$ and different spot-only line depths for a Jupiter-sized system.}
    \label{fig:jupiter_abs_spec_mean_T23_sp125_P5_shallowline}
\end{figure}

\begin{figure}[!htbp]
    \centering
    \includegraphics[width=\linewidth]{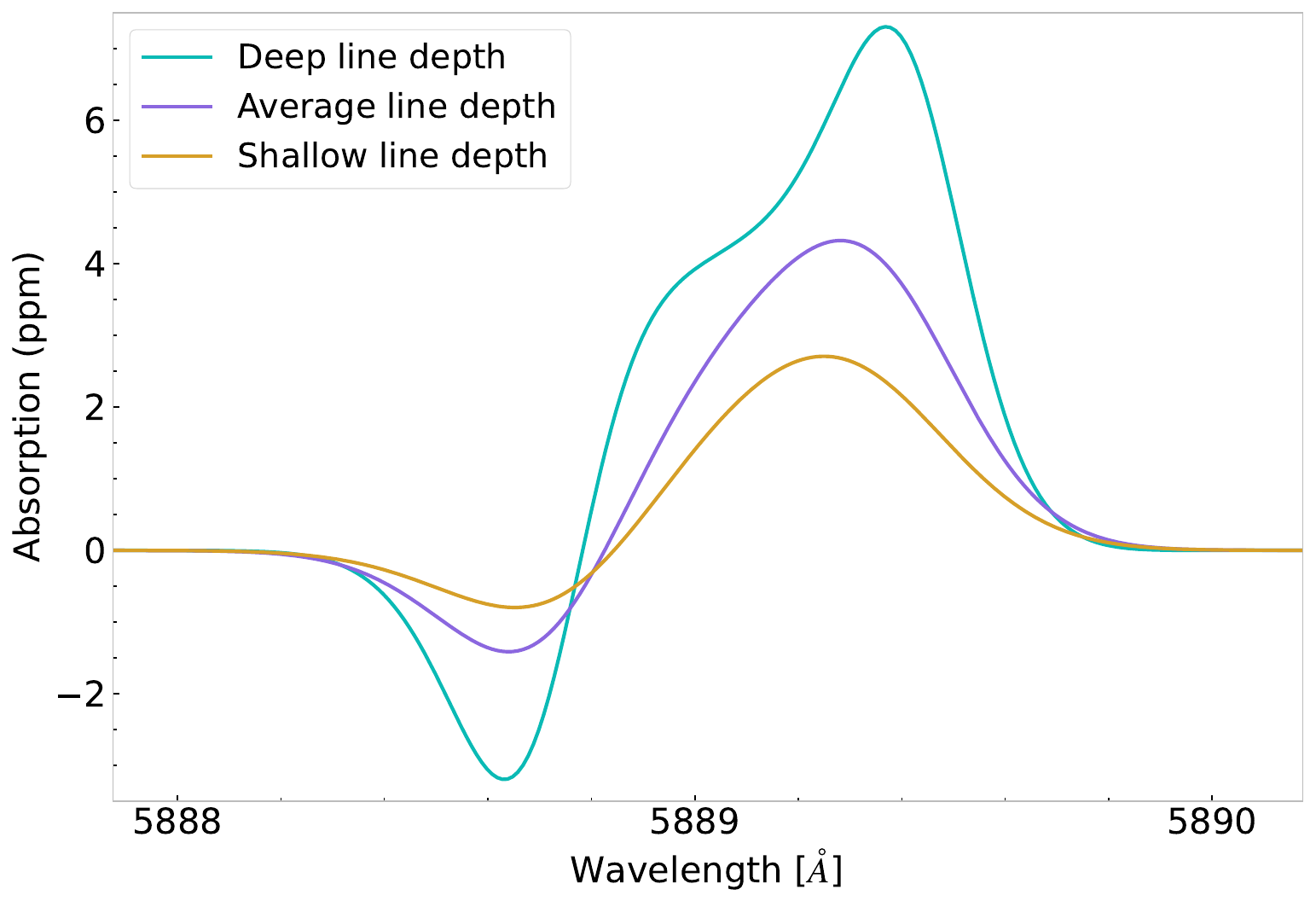}
    \caption{Absorption spectrum averaged between T$_2$ and T$_3$ in the planet's rest frame for the case with $\rm s_p = 1.25$, $\rm P_* = 5$ and different spot-only line depths for an Earth-sized system.}
    \label{fig:earth_abs_spec_mean_T23_sp125_P5_shallowline}
\end{figure}

As the choice of the line depth was arbitrary, we decided to run new simulations by changing the depth of the spot-only line. We used two different depths, which we arbitrarily named the ‘average line depth’ and the ‘shallow line depth’ (see Figure \ref{fig:spectrum}. To single out the effect of line depth, we also adjusted the width of the lines so that the profiles for the different chosen depths would cover more or less the same wavelengths as for the deep line. This allows the distortions to also cover more or less the same wavelength range (see Figs \ref{fig:jupiter_abs_spec_mean_T23_sp125_P5_shallowline} and \ref{fig:earth_abs_spec_mean_T23_sp125_P5_shallowline}). The depths of these new lines were chosen so that the depth of the average line and the depth of the shallow line would be approximately two-thirds and one-third of the initial spot-only line, respectively. The rest of the parameters were set as in our most extreme case, i.e. a spot size of 1.25 times the planetary radius and a stellar rotation period of 5 days. Decreasing the line depth of the spot-only line leads to a decrease in the amplitude of the line distortion in the absorption spectrum. This is due to a combination of the lines from the spot being shallower in the disc-integrated spectra, and the amplitude of the signal in the difference ($\rm F_{abs}$) decreasing with decreasing line depth. 


\FloatBarrier
\section{Spot temperature variation}
\label{appendix:temperature}
\begin{figure}[!htbp]
    \centering
    \includegraphics[width=\linewidth]{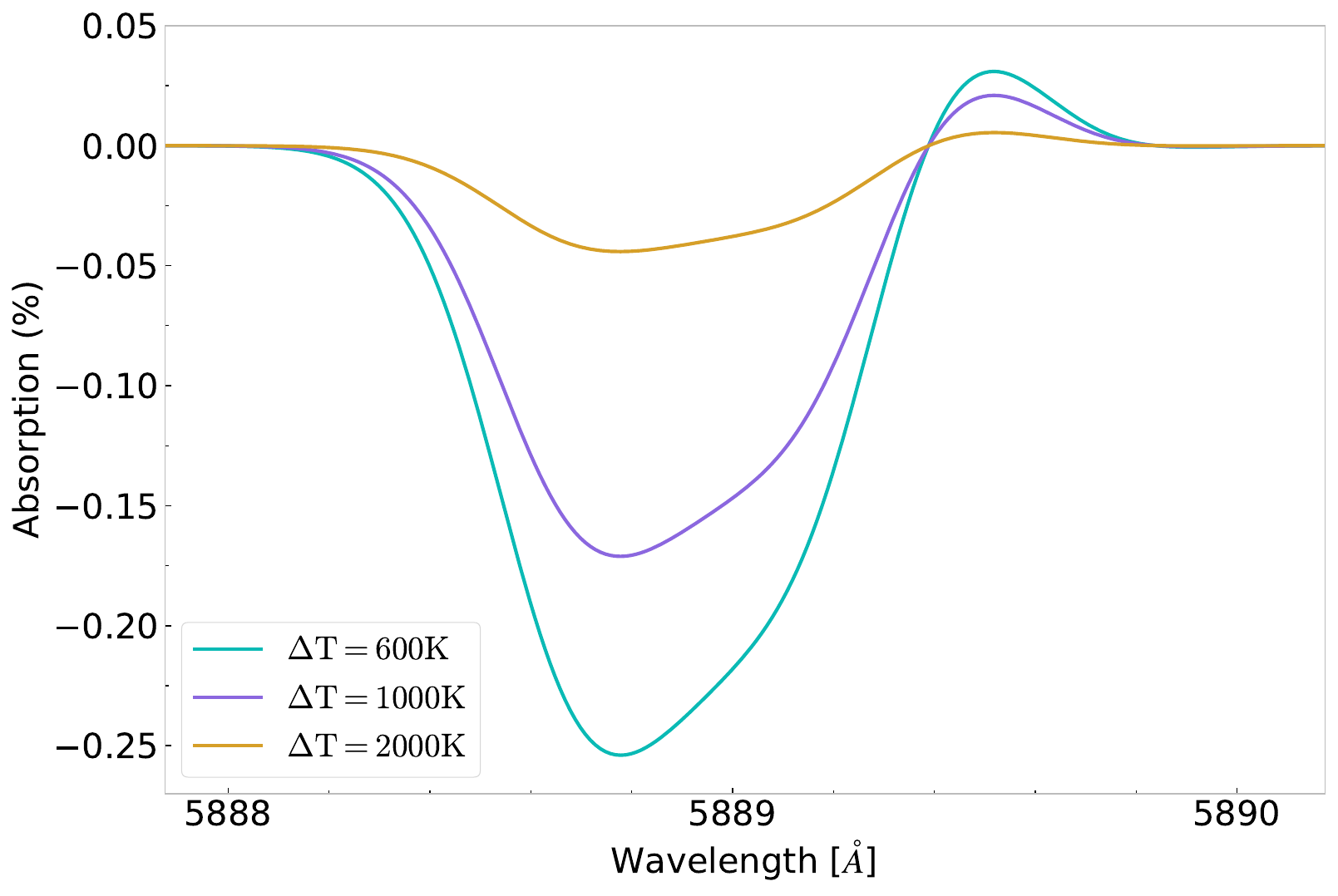}
    \caption{Absorption spectrum averaged between T$_2$ and T$_3$ in the planet's rest frame for the case with $\rm s_p = 1.25$, $\rm P_* = 5$ and different spot temperatures for a Jupiter-sized system.}
    \label{fig:jupiter_abs_spec_mean_T23_sp125_P5_T}
\end{figure}

\begin{figure}[!htbp]
    \centering
    \includegraphics[width=\linewidth]{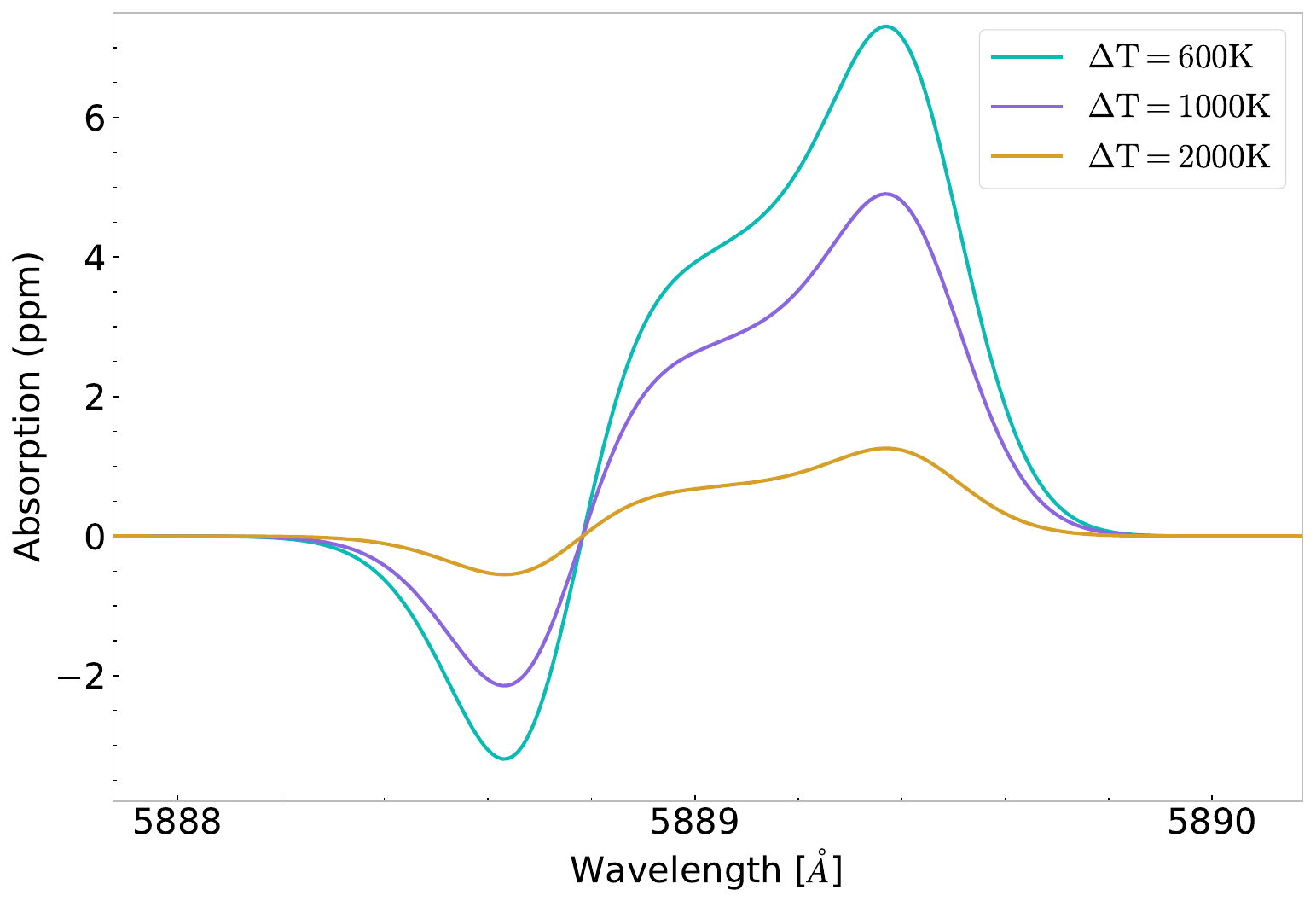}
    \caption{Absorption spectrum averaged between T$_2$ and T$_3$ in the planet's rest frame for the case with $\rm s_p = 1.25$, $\rm P_* = 5$ and different spot temperatures for an Earth-sized system.}
    \label{fig:earth_abs_spec_mean_T23_sp125_P5_T}
\end{figure}

Depending on the target, the temperature contrast between spots and the photosphere can be different from the initial value we used in the study \citep[e.g.][]{berdyugina2005,fontenla2006}.
We thus changed the temperature contrast in the simulation from 600K to 1000K and 2000K, which are arbitrarily chosen, but are of the same order of magnitude as in \citet{berdyugina2005,fontenla2006}. The rest of the parameters were set as in our most extreme case, i.e. a spot size of 1.25 times the planetary radius and a stellar rotation period of 5 days. Figure \ref{fig:jupiter_abs_spec_mean_T23_sp125_P5_T} and \ref{fig:earth_abs_spec_mean_T23_sp125_P5_T} show the mean absorption spectrum between T$_2$ and T$_3$ for the Jupiter- and Earth-sized cases, respectively. We show that increasing the temperature difference between the spot and the photosphere leads to a decrease in the amplitude of the line distortions created in the absorption spectrum. As we do not modify the depth of the spot-only line in the input spectrum, this is due to the spectrum of the spot contributing less to the disc-integrated spectra because of the lower flux level of the spectrum of the spot. We also see that the distortions keep relatively the same shape for the different cases.

\FloatBarrier
\section{Change in the wavelength range}
\label{appendix:wavelength}
One can question whether the choice of wavelength range could impact the results and conclusions of this work. For us to test the consequence of working in the optical wavelength range against, for example, the infrared wavelength range, we ran new simulations using the same spectrum as for the optical but defined on a wavelength range corresponding to the infrared. The rest of the setup was kept the same, so the only difference is the wavelength vector. This has an impact on the value of the Doppler shift when we shift the local stellar spectra to account for the stellar rotation. In the infrared, the Doppler shift is larger for the same RV and, although we use the same local line profiles, the line profile in the disc-integrated spectrum differs between the optical and the infrared. A different shape in the disc-integrated line profiles leads to differences in the POLDs of the absorption spectra.

\begin{figure}[!htbp]
    \centering
    \includegraphics[width=\linewidth]{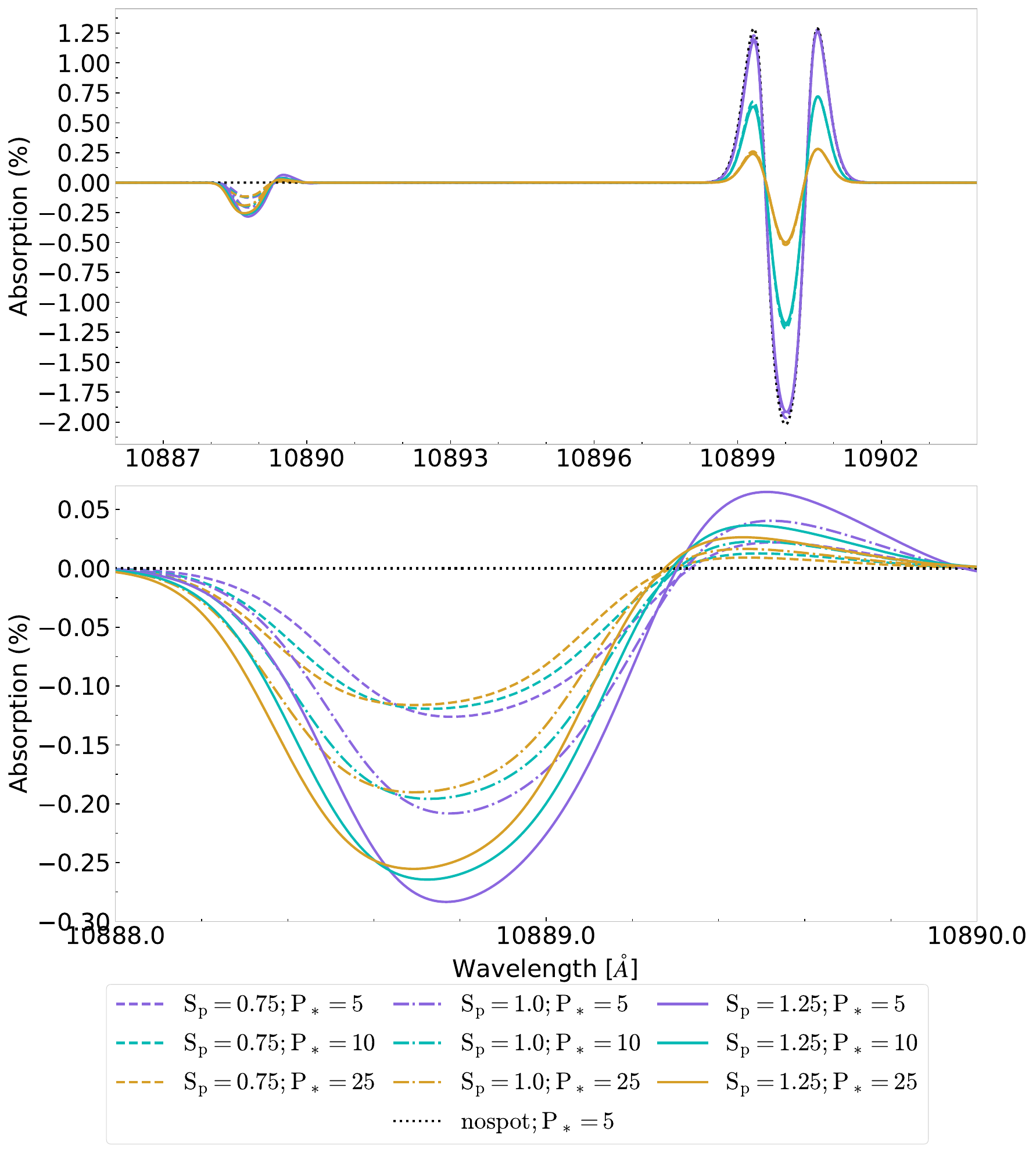}
    \caption{Upper panel: Absorption spectrum averaged between T$_2$ and T$_3$ in the planet's rest frame for the different cases for the Jupiter-sized system. Lower panel: Zoom on the wavelength range around the spot spectral line.}
    \label{fig:jupiter_IR}
\end{figure}

\begin{figure}[!htbp]
    \centering
    \includegraphics[width=\linewidth]{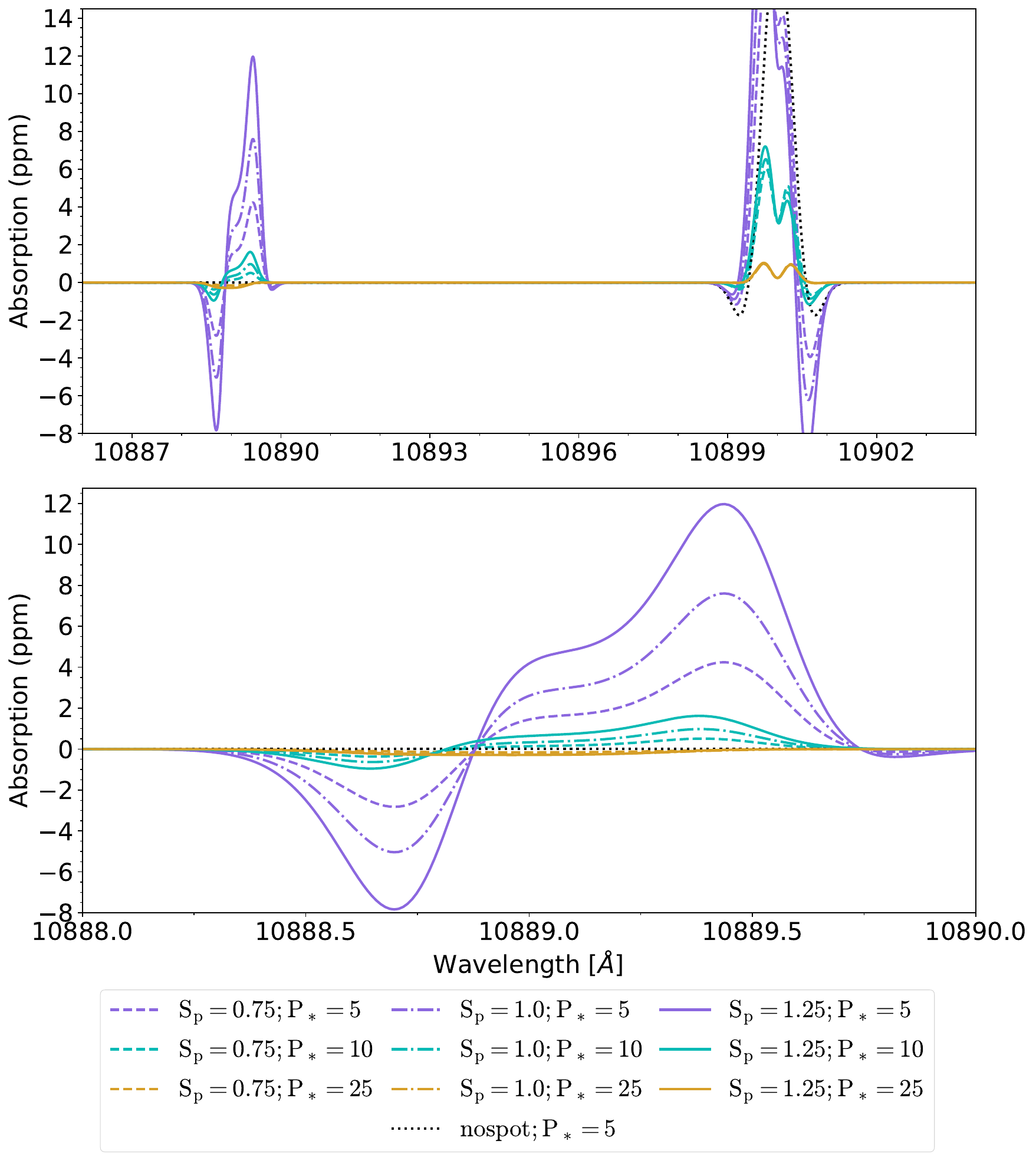}
    \caption{Upper panel: Absorption spectrum averaged between T$_2$ and T$_3$ in the planet's rest frame for the different cases for the Earth-sized system. Lower panel: Zoom on the wavelength range around the spot spectral line.}
    \label{fig:earth_IR} 
\end{figure}

 We see that the exact shapes of the distortions around the spot-only line are a bit different than for the optical wavelength range and that the amplitude of the distortions is a bit larger (see Sect. \ref{sec:result}). However, the distortions around the spot-only line are still taking place (the effect we try to highlight in this study) and are roughly similar to when working on the optical wavelength range. We thus conclude that the phenomenon that creates distortions around a spot-only line can still take place regardless of whether we are in the optical or infrared wavelength range.
\FloatBarrier
\end{appendix}
\end{document}